\documentclass[reprint,
 amsmath,amssymb,
 aps,pra, 
 showkeys
]{revtex4-2}

\usepackage{graphicx}
\usepackage{dcolumn}
\usepackage{bm}
\usepackage[version=4]{mhchem}
\usepackage{hyperref}

\begin{document}

\title{Extended phase-space symplectic integration for electron dynamics}

\author{Fran\c{c}ois Mauger$^{1}$, Cristel Chandre$^{2}$}
\affiliation{%
    $^{1}$Department of Physics and Astronomy, Louisiana State University,
        Baton Rouge, Louisiana 70803, USA \\ 
    $^{2}$CNRS, Aix Marseille Univ, I2M, Marseille, France \
}%

\date{\today}

\begin{abstract}
    We investigate the use of extended phase-space symplectic integration for simulating two different classes of electron dynamics. 
    The first one, with one and a half degrees of freedom, comes from plasma physics and describes the classical dynamics of a charged particle in a strong, constant, and uniform magnetic field perturbed by a turbulent electrostatic potential.
    The second one, with an infinite number of degrees of freedom, comes from physical chemistry and corresponds to Kohn-Sham time-dependent density-functional theory.
    For both we lay out the extension procedure and stability condition for numerical integration of the dynamics using high-order symplectic split-operator schemes. We also identify a computationally inexpensive metric that can be used for on-the-fly estimation of the accuracy of simulations.
    Our work paves the way for broad application of symplectic split-operator integration of classical and quantum Hamiltonian systems with finite and infinite number of degrees of freedom by comparing different modes of implementation of extended phase space integration.
\end{abstract}


\maketitle

\section{Introduction} \label{sec:Introduction}

Hamiltonian systems play a fundamental role across the physical and chemical sciences. They provide a unifying framework for understanding the classical, semiclassical and quantum dynamics of systems governed by energy conservation and symplectic structure. Such systems appear in a remarkably broad range of disciplines — from plasma physics and fluid dynamics~\cite{Morrison_1998} to atomic, molecular, and optical science~\cite{Mauger2009,Dubois2018}, as well as condensed matter physics~\cite{Ashcroft1976,Altland_2010}, physical chemistry~\cite{Miller1998,Gomez_Pueyo_2018,Mauger_2023}, and the study of nonlinear dynamics and chaos~\cite{Tabor1989, Masoliver2011}. In each of these fields, the Hamiltonian formulation offers powerful analytical and computational tools to describe how complex systems evolve over time, capture conserved quantities, and reveal the underlying geometrical structure of their motion~\cite{Arnold1989,Hairer2006,Marsden1999,McLachlan_2022}.

Accurate and efficient numerical simulation of the dynamics of Hamiltonian systems is of fundamental importance, as these systems often exhibit complex and high-dimensional long-term behavior. Over the years, a variety of numerical integration schemes have been developed to address these challenges, including symplectic and geometric integrators that preserve the underlying Hamiltonian structure, conserve invariants, and maintain stability over long simulation times~\cite{Yoshida_1990,McLachlan_1992,Hairer2006}. Such methods are essential for faithfully capturing both qualitative and quantitative features of Hamiltonian dynamics in physics, chemistry, and engineering applications~\cite{Kraus_2017}.

Among numerical integration techniques for Hamiltonian systems, symplectic split-operator schemes occupy a particularly important place because they are explicitly designed to preserve the symplectic structure of the underlying Hamiltonian flow. For systems with more complex Hamiltonian structures—such as Poisson or Lie-Poisson systems—they can also preserve additional invariants, including Casimir invariants. Beyond their theoretical advantages, symplectic-split integrators are easy to implement and highly versatile, allowing application across a wide range of problems from molecular dynamics and plasma physics to celestial mechanics~\cite{Laskar2001,Petit2019,Blanes_2013,Mauger_2023}. Consequently, they often outperform general-purpose integration methods, particularly in scenarios involving long integration times or highly oscillatory dynamics, where conventional schemes may accumulate significant numerical errors. Their combination of structural preservation, accuracy, and practical simplicity makes symplectic split-operator schemes an indispensable tool for simulating Hamiltonian systems.

Designing symplectic split-operator propagators is, however, far from trivial. Their construction has largely been confined to a restricted class of models in which (i) the Hamiltonian structure is canonical, and (ii) the Hamiltonian can be split into components that do not mix pairs of conjugate variables. These restrictions simplify the implementation of symplectic integration~\cite{Yoshida_1990} but also limit the applicability of standard split-operator schemes to more general or complex Hamiltonian systems. The extension of split methods to non-separable variables is possible provided that each term in the split is explicitly integrable~\cite{McLachlan_2002}. 

In this paper, we explore the extended phase-space method introduced in Refs.~\cite{Pihajoki2015,Tao_2016}, which relaxes condition~(ii) by allowing Hamiltonians that mix pairs of conjugate variables in a general setting where the split is not possible or not obvious. Furthermore, we generalize condition~(i) to encompass Hamiltonian systems with constant Poisson brackets, \textit{i.e.}, Poisson brackets that do not explicitly depend on the phase-space variables~\cite{Viscondi_2017}. This extension enables the immediate application of symplectic split-operator integration to a broader class of systems beyond the canonical setting.

To demonstrate the generality of the extended phase-space symplectic integration, we consider its application to two distinct models: a classical system with a reduced number of degrees of freedom, relevant to plasma physics, and a quantum system with an infinite number of degrees of freedom, arising in physical chemistry.
Specifically, the first model describes the classical dynamics of a charged particle in a strong, constant, and uniform magnetic field, perturbed by a turbulent electrostatic potential. In this system, the conjugate variables are strongly coupled in the Hamiltonian, making it unsuitable for conventional symplectic split-operator schemes.
The second model corresponds to Kohn-Sham time-dependent density-functional theory (TDDFT)~\cite{Kohn_1965,Runge_1984}, and is widely used to describe electron dynamics in atomic and molecular systems. In~\cite{Mauger_2023}, we showed that, with restrictions on the domain discretization (grid based) and DFT functional (local-density approximation, LDA, or generalized-gradient approximation, GGA), one can use symplectic split-operator akin to those for separable Hamiltonian systems with a finite number of degrees of freedom. However, the use of hybrid functionals or a basis set discretization, as is common in high-accuracy quantum-chemistry calculations, remained out of reach. We remove those limitations with the extended phase-space symplectic integration and lay out general-purpose high-order integration schemes for TDDFT dynamics.
We also leverage our two examples to explore stability conditions for the phase-space extension and identify an easy-to-calculate metric for estimating the accuracy of numerical results.

The paper is organized as follows:
Section~\ref{sec:Methods} lays out the extended phase space method, how to use it in simulations with symplectic split-operator schemes, a discussion on the stability of the extended phase-space dynamics, and the extension of the method to non-canonical Hamiltonians systems with a constant Poisson bracket.
Section~\ref{sec:ExB} discusses our first application example, with the one
and a half degrees of freedom (DOFs) classical dynamics of a charged particle in a strong, constant, and uniform magnetic field, subjected to a turbulent electrostatic potential. The Python codes for our simulation examples are available on their GitHub repository~\cite{GC2D}.
Section~\ref{sec:TDDFT} discusses our second application example, with the infinite number of DOFs dynamics associated with Kohn-Sham (KS) TDDFT. We detail the adjustment that must be made to the phase-space extension due to the complex-valued equations of motion and stabilization of numerical simulations when implementing the propagation on a grid. The TDDFT propagation schemes we use to illustrate our results are available as part of the \texttt{QMol-grid} package~\cite{mauger_2024} and accessible on the package GitHub and MathWorks repositories~\cite{QMol_grid}.
Finally, section~\ref{sec:Conclusion} concludes our work.

\section{Methods} \label{sec:Methods}

In this section, we lay out the phase-space extension procedure, including when the system is externally driven via an explicit time dependence in the Hamiltonian, and its use for symplectic integration using split-operator schemes. We also discuss the stability properties of the extended phase-space dynamics when a restraint is used, which ultimately determines the suitability of the method for numerical simulations.

\subsection{Extended phase-space integration} \label{sec:Methods:Extended_phase_space}

We consider a canonical Hamiltonian system defined by the Hamiltonian $H({\bf q},{\bf p})$, where ${\bf q}=(q_1,q_2,\ldots)$ and ${\bf p}=(p_1,p_2,\ldots)$ are the canonically conjugate variables, so that the canonical Poisson bracket is
\begin{equation} \label{eq:canonical_Poisson_bracket}
    \{F,G\} = \sum_k{
                \frac{\partial F}{\partial p_k}\frac{\partial G}{\partial q_k} - 
                \frac{\partial F}{\partial q_k}\frac{\partial G}{\partial p_k}
                }.
\end{equation}
For systems with an infinite number of degrees of freedom, the sum and partial derivatives in the Poisson bracket are replaced with an integral and functional derivatives with respect to the field ${\bf p}$ and ${\bf q}$, respectively.
The dynamics for any observable $F({\bf q},{\bf p})$ is given by
\begin{equation}  \label{eq:Hamiltonian_flow}
    \dot{F} = \mathcal{L}_{H} F = \{H,F\},
\end{equation}
where the overhead dot denotes the temporal derivative, $\mathcal{L}_{H} = \{H,\cdot\}$ is the Liouville operator and from which we recover Hamilton's equations by taking $F=q_k$ and $F=p_k$. The formal evolution of the value of any function $F$ of the phase-space variables from $t$ to $t+\tau$ is given by
\begin{equation*}
    F(t+\tau) = \text{e}^{\tau \mathcal{L}_{H}} F(t).
\end{equation*}
Integrating a Hamiltonian system amounts to computing the exponential of the Liouville operator, which is a very delicate task~\cite{Moler_2003}.
When the Hamiltonian splits between two parts that solely depend on one of the conjugate variables $H({\bf q},{\bf p})=H_1({\bf q})+H_2({\bf p})$, the dynamics can easily be integrated numerically using symplectic-split operators techniques given that the flows generated by the $H_1$ and $H_2$ pieces individually are analytically solvable
\begin{align*}
     &\text{e}^{\tau \mathcal{L}_{H_1}}{\bf q} = {\bf q} & \text{and}\quad 
     &\text{e}^{\tau \mathcal{L}_{H_1}}{\bf p} = {\bf p} - \tau \frac{\partial H_1}{\partial {\bf q}}, \\
     &\text{e}^{\tau \mathcal{L}_{H_2}}{\bf q} = {\bf q} + \tau \frac{\partial H_2}{\partial {\bf p}} &\text{and}\quad 
     &\text{e}^{\tau \mathcal{L}_{H_2}}{\bf p} = {\bf p} .
\end{align*}
In the general case where no such decomposition is doable, using a symplectic integration is less straightforward. 

In Refs.~\cite{Pihajoki2015,Tao_2016}, a general-purpose avenue to leveraging symplectic-split integration was proposed by extending the phase $({\bf q},{\bf p})\to({\bf q},{\bf p},\overline{\bf q},\overline{\bf p})$ together with an extended Hamiltonian with a restraint term
\begin{eqnarray}
    \overline{H}({\bf q},{\bf p},\overline{\bf q},\overline{\bf p}) & = & 
        H({\bf q},\overline{\bf p}) + H(\overline{\bf q},{\bf p}) \nonumber \\ && + \frac{\omega}{2} \left(
                \left\|{\bf q}-\overline{\bf q}\right\|^2 + 
                \left\|{\bf p}-\overline{\bf p}\right\|^2
            \right), \label{eq:extended_Hamiltonian}
\end{eqnarray}
for some properly chosen restraint coefficient $\omega\geq0$, and the canonical extended Poisson bracket acting on functions $\overline{F}({\bf q},{\bf p},\overline{\bf q},\overline{\bf p})$,
\begin{equation} \label{eq:extended_Poisson_bracket}
    {\{\overline{F},\overline{G}\}} = \sum_k{
                \frac{\partial \overline{F}}{\partial p_k}\frac{\partial \overline{G}}{\partial q_k} - 
                \frac{\partial \overline{F}}{\partial q_k}\frac{\partial \overline{G}}{\partial p_k} +
                \frac{\partial \overline{F}}{\partial \overline{p}_k}\frac{\partial \overline{G}}{\partial \overline{q}_k} - 
                \frac{\partial \overline{F}}{\partial \overline{q}_k}\frac{\partial \overline{G}}{\partial \overline{p}_k}
                }.
\end{equation}
We notice the uncoupling of the pairs of canonically conjugate variables in each instance of the Hamiltonian $H$. The extended phase-space dynamics is then governed by the equation of motion
\begin{eqnarray*}
    &&\dot{\bf q} = {\mathcal{L}}_{\overline{H}} {\bf q} = 
        {\{\overline{H},{\bf q}\}} = 
        +\left. \frac{\partial H}{\partial {\bf p}}\right|_{\overline{\bf q},{\bf p}} + 
        \omega ({\bf p} - \overline{\bf p}),\\
    &&\dot{\bf p} = {\mathcal{L}}_{\overline{H}} {\bf p} = 
        {\{\overline{H},{\bf p}\}} = 
        -\left. \frac{\partial H}{\partial {\bf q}}\right|_{{\bf q},\overline{\bf p}} - 
        \omega ({\bf q} - \overline{\bf q}),
\end{eqnarray*}
and similarly for the $\overline{\bf q}$ and $\overline{\bf p}$ variables. We then see that any extended phase-space trajectory that maintains $\overline{\bf q}={\bf q}$ and $\overline{\bf p}={\bf p}$ is a solution of the original (non-extended) system. It constitutes an invariant submanifold.

For numerical simulations, the phase-space extension of Eqs.~\eqref{eq:extended_Hamiltonian}-\eqref{eq:extended_Poisson_bracket} doubles the number of equations of motion to integrate. Yet, it defines three pieces in the extended Hamiltonian $H_1=H({\bf q},\overline{\bf p})$, $H_2=H(\overline{\bf q},{\bf p})$, and the restraint $R({\bf q},{\bf p},\overline{\bf q},\overline{\bf p})$ that individually generate analytically integrable flows and thus is compatible with symplectic split-operator schemes~\cite{Hairer2006,McLachlan_2022}, as we discuss in the following section. For $H_1$, we get
\begin{subequations} \label{eq:extended_split_Hamiltonian_flow}
\begin{align}
     &\text{e}^{\tau {\mathcal{L}}_{H_1}}{\bf q} = {\bf q} 
        & \text{and}\quad 
        &\text{e}^{\tau {\mathcal{L}}_{H_1}}{\bf p} = {\bf p} - 
            \tau \frac{\partial H_1}{\partial {\bf q}}, \\
     &\text{e}^{\tau {\mathcal{L}}_{H_1}}\overline{\bf q} = \overline{\bf q} + 
            \tau \frac{\partial H_1}{\partial \overline{\bf p}} 
        &\text{and}\quad 
        &\text{e}^{\tau {\mathcal{L}}_{H_1}}\overline{\bf p} = \overline{\bf p},
\end{align}
\end{subequations}
given that $\frac{\partial H_1}{\partial {\bf p}}=\frac{\partial H_1}{\partial \overline{\bf q}}=0$, and similarly for the flow generated by $H_2$. 
In fact, $\text{e}^{\tau {\mathcal{L}}_{H_{1,2}}}$ can be seen as the linearization of the original (non-extended) Hamiltonian flow in the mixed coordinate spaces $({\bf q},\overline{\bf p})$ and $(\overline{\bf q},{\bf p})$.
For the restraint
$
    R({\bf q},{\bf p},\overline{\bf q},\overline{\bf p}) = 
        \omega \left(
                    \left\|{\bf q}-\overline{\bf q}\right\|^2 + 
                    \left\|{\bf p}-\overline{\bf p}\right\|^2
                \right)/2,
$
after integration of its equations of motion, we get
\begin{subequations} \label{eq:extended_restrain_flow}
\begin{align}
    & \text{e}^{\tau {\mathcal{L}}_{R}}{\bf q} = 
        \frac{{\bf q}+\overline{\bf q}}{2} + 
        \cos(2\omega\tau) \frac{{\bf q}-\overline{\bf q}}{2} + 
        \sin(2\omega\tau) \frac{{\bf p}-\overline{\bf p}}{2}, \\
    & \text{e}^{\tau {\mathcal{L}}_{R}}{\bf p} = 
        \frac{{\bf p}+\overline{\bf p}}{2} - 
        \sin(2\omega\tau) \frac{{\bf q}-\overline{\bf q}}{2} + 
        \cos(2\omega\tau) \frac{{\bf p}-\overline{\bf p}}{2},
\end{align}
\end{subequations}
and changing the signs in front of the trigonometric functions for $\overline{\bf q}$ and $\overline{\bf p}$.
Geometrically, the flow associated with the restraint corresponds to a rotation of the separation vector between the pairs of extended coordinates $\delta{\bf q}={\bf q}-\overline{\bf q}$ and $\delta{\bf p}={\bf p}-\overline{\bf p}$ around the average trajectory $(({\bf q}+\overline{\bf q})/2,({\bf p}+\overline{\bf p})/ 2)$.

For externally driven systems, the Hamiltonian includes an explicit time dependence $H({\bf q},{\bf p},t)$.
To autonomize the extended Hamiltonian system, we consider a synchronous approach where the two Hamiltonian copies in Eq.~\eqref{eq:extended_Hamiltonian} share the same time and we add a single variable $\xi$ canonically conjugated to $t$
\begin{eqnarray}
    \overline{H}_{\rm a}({\bf q},{\bf p},\overline{\bf q},\overline{\bf p},t,\xi) & = & 
        H({\bf q},\overline{\bf p},t) + H(\overline{\bf q},{\bf p},t) + \xi \nonumber \\ 
    && + \frac{\omega}{2} \left(
                \left\|{\bf q}-\overline{\bf q}\right\|^2 + 
                \left\|{\bf p}-\overline{\bf p}\right\|^2
            \right). \label{eq:autonomized_extended_Hamiltonian}
\end{eqnarray}
The autonomized extended Poisson bracket acting on functions $\overline{F}({\bf q},{\bf p},\overline{\bf q},\overline{\bf p},t,\xi)$ then reads
\begin{eqnarray}
    && {\{\overline{F},\overline{G}\}} = \sum_k{
                \frac{\partial \overline{F}}{\partial p_k}\frac{\partial \overline{G}}{\partial q_k} - 
                \frac{\partial \overline{F}}{\partial q_k}\frac{\partial \overline{G}}{\partial p_k} +
                \frac{\partial \overline{F}}{\partial \overline{p}_k}\frac{\partial \overline{G}}{\partial \overline{q}_k} - 
                \frac{\partial \overline{F}}{\partial \overline{q}_k}\frac{\partial \overline{G}}{\partial \overline{p}_k}
                } \nonumber \\ && \qquad \qquad + 
            \frac{\partial \overline{F}}{\partial \xi}\frac{\partial \overline{G}}{\partial t} - 
            \frac{\partial \overline{F}}{\partial t}\frac{\partial \overline{G}}{\partial \xi}.\label{eq:autonomized_extended_Poisson_bracket}
\end{eqnarray}
Since the autonomized Hamiltonian $\overline{H}_{\rm a}$ is conserved, it can be used as an estimator for the accuracy of numerical simulations.

\subsection{Symplectic split-operator propagation schemes} \label{sec:Methods:symplectic_schemes}

After applying the extended phase space procedure described in the previous section, the Hamiltonian can be decomposed as
\begin{equation} \label{eq:Hamiltonian_split}
    H = H_1 + H_2 + \ldots + H_m,
\end{equation}
where each $\text{e}^{\tau \mathcal{L}_{H_k}}$ is easily and accurately computable. Specifically, for Hamiltonian~\eqref{eq:extended_Hamiltonian}, we have $H_1 = H({\bf q},\overline{\bf p})$, $H_2 = H(\overline{\bf q},{\bf p})$, and, in case of the use of a restraint, $H_3=R({\bf q},{\bf p},\overline{\bf q},\overline{\bf p})$. 
When considering the application to TDDFT, we will further split the Hamiltonian in additional components and thus keep the discussion here with an arbitrary number $m>2$.
For the autonomized Hamiltonian~\eqref{eq:autonomized_extended_Hamiltonian}, we group the variable $\xi$ with the restraint such that $H_3=R+\xi$. For $\omega=0$, like in the midpoint projection discussed below, $H_3$ reduces to $\xi$ only. Then, in addition to Eq.~\eqref{eq:extended_restrain_flow}, the flow generated by $H_3$ includes
\begin{equation}
    \text{e}^{\tau\mathcal{L}_{H_3}} t = t + \tau \quad \text{and} \quad
    \text{e}^{\tau\mathcal{L}_{H_3}} \xi = \xi,
\end{equation}
which indicates that time should be updated after each application of $\text{e}^{\tau\mathcal{L}_{H_3}}$ in split-operator schemes.
If one is interested in tracking the conservation of the Hamiltonian during the propagation, in addition to Eqs.~\eqref{eq:extended_restrain_flow}, the flow generated by $H_1$ includes
\begin{equation} \label{eq:extended_split_Hamiltonian_autonomized_flow}
    \text{e}^{\tau\mathcal{L}_{H_1}} t = t \quad \text{and} \quad
    \text{e}^{\tau\mathcal{L}_{H_1}} \xi = \xi - \tau \left.\frac{\partial H_1}{\partial t}\right|_{{\bf q},\overline{\bf p}},
\end{equation}
given that ${\bf q}$ and $\overline{\bf p}$ are constant, and similarly for the flow generated by $H_2$.

Considering the split of Eq.~\eqref{eq:Hamiltonian_split} and following~\cite{McLachlan_2022}, we use standard split-operator schemes of the form
\begin{equation} \label{eq:symplectic_split_operator}
    {\rm e}^{\tau \mathcal{L}_H} = \prod_{k=1}^K{
        \chi_{\alpha_{2k}\tau} \ \chi^*_{\alpha_{(2k-1)}\tau}}
        + \mathcal{O}\left(\tau^{p+1}\right),
\end{equation}
where $p$ is the order of the scheme and $K$ the number of steps, and
\begin{equation} \label{eq:symplectic_chi_fb}
    \chi_\tau   = \prod_{k=1}^{m}{{\rm e}^{\tau \mathcal{L}_{H_{k}}}}
    \quad \text{and} \quad
    \chi^*_\tau = \prod_{k=1}^{m}{{\rm e}^{\tau \mathcal{L}_{H_{m+1-k}}}},
\end{equation}
for which we have analytical solutions for the extended phase space variables, as defined in the various equations above. In addition, we consider time reversible split schemes, \textit{i.e.}, schemes such that $\alpha_{2K+1 -k}=\alpha_{k}$ for $k=1,\ldots,K$.
Here we take the convention that the product of operators above apply from right to left. For instance, Eq.~\eqref{eq:symplectic_split_operator} first applies $\chi^*_{\alpha_1\tau}$, followed by $\chi_{\alpha_2\tau}$, $\chi^* _{\alpha_3\tau}$, \ldots, $\chi_{\alpha_{2K}\tau}$.
Note that the order of the exponentials is reversed between $\chi_\tau$ and $\chi^*_\tau$, such that the last element in one is always the first in the other, or in other terms $\chi_\tau^*=(\chi_{-\tau})^{-1}$. These pairs of last-first exponentials can thus be refactored 
$
    \text{e}^{\alpha \tau \mathcal{L}_{H_k}}
    \text{e}^{\alpha^\prime \tau \mathcal{L}_{H_k}} =
    \text{e}^{\left(\alpha+\alpha^\prime \right) \tau \mathcal{L}_{H_k}},
$  
when evaluating Eq.~\eqref{eq:symplectic_split_operator} to reduce the computational footprint of the split operator scheme~\cite{McLachlan_2022}.
Since, for arbitrary $\alpha$, $\tau$, and $k$, each of the ${\rm e}^{\alpha\tau \mathcal{L}_{H_k}}$ is a canonical transformation this ensures that the overall scheme is symplectic in the extended phase space.
The order of the scheme is then defined by judiciously chosen sets of coefficients $\{a_k\}_k$. To keep figures legible, in the illustration cases of Secs.~\ref{sec:ExB} and~\ref{sec:TDDFT} we use the optimized 4$^\text{th}$-order Blanes and Moan scheme~\cite{Blanes_2002}, which provides a good balance between accuracy and efficacy~\cite{McLachlan_2022}. For this scheme, we have the order $p=4$ and a number of steps $K=6$. We have observed qualitatively similar results with other split-operators schemes -- see also Appendix~\ref{app:split_scheme}.

The split operator~\eqref{eq:symplectic_split_operator} provides a workable algorithm to integrate the extended phase-space dynamics. 
Starting from the initial condition $\overline{\bf q}_0={\bf q}_0$ and $\overline{\bf p}_0={\bf p}_0$, the split operator flow only preserves the symmetry between the pairs of extended phase space variables at the order of the scheme \emph{and provided the extended dynamics is stable} -- see also the discussion in the following section.
Yet, in practice, we are interested in a numerical solution in the original (non-extended) phase space. 
Throughout this article, we approximate this solution as the extended phase-space average $\left(\frac{{\bf q}+\overline{\bf q}}{2},\frac{{\bf p}+\overline{\bf p}}{2}\right)$ once the integration in the extended phase space is completed. For consistency, we likewise use the average to calculate all observable $F\left(\frac{{\bf q}+\overline{\bf q}}{2},\frac{{\bf p}+\overline{\bf p}}{2}\right)$.
Interestingly, when using the restraint, we have observed that the distance between the pairs of extended phase space variables $\sqrt{\|{\bf q}-\overline{\bf q}\|^2+\|{\bf p}-\overline{\bf p}\|^2}$ strongly correlates with the accuracy of the average trajectory and associated observables. It therefore provides a computationally inexpensive metric for the accuracy of simulations for the extended integration with a restraint.
Finally, for externally driven systems, we note that the variable $\xi$ gets updated in both $\text{e}^{\alpha \tau\mathcal{L}_{H_1}}$ and $\text{e}^{\alpha\tau\mathcal{L}_{H_2}}$, reflecting the two base-Hamiltonian components in the extension of Eq.~\eqref{eq:autonomized_extended_Hamiltonian}. In turn, it means that conservation of the autonomized Hamiltonian should be checked as defined by 
\begin{equation} \label{eq:energy_conservation}
    H\left(\frac{{\bf q}+\overline{\bf q}}{2},\frac{{\bf p}+\overline{\bf p}}{2}\right)+\frac{\xi}{2}.
\end{equation}

\subsection{Influence of the restraint on the extended phase-space dynamics} \label{sec:Methods:restrain_influence}

For numerical simulations, a key parameter is the choice of the restraint coefficient $\omega$ in the extended Hamiltonian~\eqref{eq:extended_Hamiltonian} such that the average trajectory $(\frac{{\bf q}+\overline{\bf q}}{2},\frac{{\bf p}+\overline{\bf p}}{2})$ matches the original (non-extended) Hamiltonian flow, at the order of the integration scheme. 
To build some intuition on the role of $\omega$, we consider a general Hamiltonian system with a single degree of freedom $H(q,p)$, where $q$ and $p$ are scalar variables.
We apply the phase-space extension~\eqref{eq:extended_Hamiltonian}, followed by the canonical change of coordinates
$$
    \tilde{q} = \frac{q+\overline{q}}{\sqrt{2}}, \quad
    \delta\tilde{q} = \frac{q-\overline{q}}{\sqrt{2}}, \quad
    \tilde{p} = \frac{p+\overline{p}}{\sqrt{2}}, \quad 
    \delta\tilde{p} = \frac{p-\overline{p}}{\sqrt{2}},
$$
such that the extended Hamiltonian reads
\begin{eqnarray}
    \overline{H}(\tilde{q},\delta\tilde{q},\tilde{p},\delta\tilde{p}) & = & 
        H\left(
            \frac{\tilde{q}+\delta\tilde{q}}{\sqrt{2}},
            \frac{\tilde{p}-\delta\tilde{p}}{\sqrt{2}}
        \right) \nonumber \\ &&  
        + H\left(
            \frac{\tilde{q}-\delta\tilde{q}}{\sqrt{2}},
            \frac{\tilde{p}+\delta\tilde{p}}{\sqrt{2}}
        \right) \nonumber \\ &&
        + \omega ( \delta\tilde{q}^2 + \delta\tilde{p}^2), \label{eq:extended_Hamiltonian_example}
\end{eqnarray}
with the Poisson bracket
$$
    {\{\overline{F},\overline{G}\}} = 
            \frac{\partial \overline{F}}{\partial \tilde{p}}\frac{\partial \overline{G}}{\partial \tilde{q}} - 
            \frac{\partial \overline{F}}{\partial \tilde{q}}\frac{\partial \overline{G}}{\partial \tilde{p}} +
            \frac{\partial \overline{F}}{\partial \delta\tilde{p}}\frac{\partial \overline{G}}{\partial \delta\tilde{q}} - 
            \frac{\partial \overline{F}}{\partial \delta\tilde{q}}\frac{\partial \overline{G}}{\partial \delta\tilde{p}}.
$$

Next, we consider the second-order Taylor expansion of the extended Hamiltonian~\eqref{eq:extended_Hamiltonian_example}, in the separation variables $\delta\tilde{q}$ and $\delta\tilde{p}$
\begin{equation} \label{eq:linearized_Hamiltonian_example}
    \tilde{H} = 2  H + \frac{1}{2}
        \left( \begin{array}{c} 
            \delta\tilde{q} \\ \delta\tilde{p}
        \end{array} \right) \cdot
        \left(\begin{array}{cc} 
            \partial^2_{qq}H + 2\omega & -\partial^2_{qp}H \\ 
            -\partial^2_{qp}H          & \partial^2_{pp}H + 2\omega
        \end{array} \right)  \left( \begin{array}{c} 
            \delta\tilde{q} \\ \delta\tilde{p}
        \end{array} \right),
\end{equation}
up to order 3 in $\delta \tilde{q}$ and $\delta \tilde{p}$. Here $H$ and its derivatives are all evaluated at the average extended phase space coordinate $\left(\frac{\tilde{q}}{\sqrt{2}},\frac{\tilde{p}}{\sqrt{2}}\right) = \left(\frac{q+\overline{q}}{2},\frac{p+\overline{p}}{2}\right)$, and we assume the Hamiltonian to be a smooth function of $q$ and $p$, such that $\partial^2_{qp}H=\partial^2_{pq}H$. Equation~\eqref{eq:linearized_Hamiltonian_example} tells us that the Hamiltonian in the original phase space is approximately conserved provided the distance between the two copies remains close to each other. 
At any point along the extended phase-space trajectory, we use Hamiltonian~\eqref{eq:linearized_Hamiltonian_example} to study the local linear stability of the dynamics in the separation variables. The (linearized) equation of motion for the separation variables become
$$
    \frac{d}{dt}\left(\begin{array}{c} \delta\tilde{q} \\ \delta\tilde{p} \end{array}\right) = 
        \left(\begin{array}{cc} 
            -\partial^2_{qp}H         & \partial^2_{pp}H+2\omega \\
            -\partial^2_{pp}H-2\omega & \partial^2_{qp}H
        \end{array}\right) \left(\begin{array}{c} \delta\tilde{q} \\ \delta\tilde{p} \end{array}\right),
$$
which can be solved analytically by diagonalizing the matrix in the right-hand term.
The characteristic polynomial for the matrix is
$$
    \lambda^2 - (\partial^2_{qp}H)^2 + (\partial^2_{pp}H+2\omega) (\partial^2_{qq}H+2\omega) = 0,
$$
such that the linearized dynamics does not exhibit exponential growth in the separation variables when the polynomial has only purely imaginary solutions or equivalently when
\begin{equation} \label{eq:example_stability_condition}
    (\partial^2_{pp}H+2\omega) (\partial^2_{qq}H+2\omega) - (\partial^2_{qp}H)^2  > 0.
\end{equation}

The stability condition~\eqref{eq:example_stability_condition} corresponds to a quadratic polynomial in $\omega$, with leading term $4\omega^2$, and thus can always be fulfilled provided $\omega$ is chosen large enough. In turn, it suggests that one may always find a restraint coefficient that avoids the divergence and thus could provide an average trajectory solution of the original Hamiltonian system, provided that each part of the split converges.
We note $\omega_1\leq\omega_2$ the two real-valued roots of the polynomial with respect to $\omega$ in the left-hand term of the stability equation~\eqref{eq:example_stability_condition} and distinguish three cases for the linearized dynamics stability condition:
(i) If $\omega_2<0$, the dynamics is unconditionally stable and any value of the restraint coefficient $\omega\geq0$ can be chosen, including not applying any restraint at all $R=0$.
(ii) If $\omega_2>0$ and $\omega_1<0$, $\omega_2$ serves as a critical stability condition above which the restraint coefficient should be chosen $\omega>\omega_2$.
(iii) Lastly, when $\omega_1>0$ we have two distinct stability regions with $0\leq\omega<\omega_1$, again allowing to ignore the restraint altogether, and $\omega>\omega_2$.
For cases (i) and (iii), we note that the stability condition fulfillment for $\omega=0$ is equivalent to saying that the dynamics for the original (non-extended) Hamiltonian system is itself locally stable. It thus makes sense that no additional restraint need be imposed on its extended counterpart.
In all cases, taking $\omega>\max(0,\omega_2)$ ensures satisfying the stability condition~\eqref{eq:example_stability_condition}. This is the condition we aim to generally satisfy in setting the default value for $\omega$ in the simulations we discuss below.

\subsection{Extension of the restraint method to constant Poisson brackets} \label{sec:Methods:non_canonical_brackets}

Formally, by exchanging the role of, \textit{e.g.}, ${\bf q}$ and $\bar{\bf q}$, the extended symplectic-split scheme is equivalent to extending the Hamiltonian in the following way
\begin{eqnarray*}
    \overline{H}({\bf q},{\bf p},\overline{\bf q},\overline{\bf p}) & = & 
        H({\bf q},{\bf p}) + H(\overline{\bf q},\overline{\bf p}) \nonumber \\ 
    && + \frac{\omega}{2} \left(
                \left\|{\bf q}-\overline{\bf q}\right\|^2 + 
                \left\|{\bf p}-\overline{\bf p}\right\|^2
            \right),
\end{eqnarray*}
and extending and weaving the Poisson bracket acting on functions $\overline{F}({\bf q},{\bf p},\overline{\bf q},\overline{\bf p})$ yields
$$
    \{\overline{F},\overline{G}\} = \frac{\partial \overline{F}}{\partial {\bf p}}\cdot \frac{\partial \overline{G}}{\partial \bar{\bf q}} - 
                \frac{\partial \overline{F}}{\partial \bar{\bf q}}\cdot \frac{\partial \overline{G}}{\partial {\bf p}} + \frac{\partial \overline{F}}{\partial \bar{\bf p}}\cdot \frac{\partial \overline{G}}{\partial {\bf q}} - 
                \frac{\partial \overline{F}}{\partial {\bf q}}\cdot \frac{\partial \overline{G}}{\partial \bar{\bf p}}.
$$
Rewriting the extension this way allows us to consider more general Poisson brackets, for instance, by considering a set of (non-canonical) variables ${\bf z}$ such that the Poisson bracket is defined by a constant and antisymmetric matrix ${\mathbb J}$, and reads 
$$
    \{F,G\} = \frac{\partial F}{\partial {\bf z}} \cdot {\mathbb J} \frac{\partial G}{\partial {\bf z}} = \sum_{k,l}{ \frac{\partial F}{\partial z_k} J_{kl} \frac{\partial G}{\partial z_l}}.
$$
Notice that since the antisymmetric matrix ${\mathbb J}$ is constant, it necessarily satisfies the Jacobi identity. To implement the extended symplectic-split scheme with a restraint, we consider
$$
    \overline{H}({\bf z}, \bar{\bf z}) = H({\bf z}) + H(\bar{\bf z}) + \frac{\omega}{2}\Vert {\bf z}-\bar{\bf z}\Vert^2,
$$
and the extended Poisson bracket
$$
    \{\overline{F},\overline{G}\} = \frac{\partial \overline{F}}{\partial {\bf z}} \cdot {\mathbb J} \frac{\partial \overline{G}}{\partial \bar{\bf z}} + \frac{\partial \overline{F}}{\partial \bar{\bf z}} \cdot {\mathbb J} \frac{\partial \overline{G}}{\partial {\bf z}}.
$$
The major change is the integration of the dynamics associated with the restraint $R({\bf z}, \bar{\bf z})$. The equations of motion for this restraint are
\begin{eqnarray*}
    && \dot{\bf z} = -\omega {\mathbb J} ({\bf z}-\bar{\bf z}),\\
    && \dot{\bar{\bf z}} = \omega {\mathbb J} ({\bf z}-\bar{\bf z}),
\end{eqnarray*}
from which we deduce that ${\bf z}+\bar{\bf z}$ is constant and $\delta {\bf z}={\bf z}-\bar{\bf z}$ evolves as
$$
\delta {\bf z}(t+\tau)= \exp(-2\omega \tau {\mathbb J}) \delta {\bf z}(t).
$$
Since ${\mathbb J}$ is antisymmetric, the exponential $\exp(-2\omega \tau {\mathbb J})$ is a rotation, and can be easily computed by block-diagonalizing ${\mathbb J}$. 

The extension to a more general Poisson matrix $\mathbb{J}({\bf z})$ raises two main difficulties: First, extending the Poisson bracket is not straightforward, since the explicit dependence of $\mathbb{J}$ on $\mathbf{z}$ may violate the Jacobi identity for the extended bracket. Second, the integration of the corresponding equations of motion for the restraint $R({\bf z}, \bar{\bf z})$ may no longer be achievable in closed form. Some attempts at generalization can be found in Ref.~\cite{Zhu_2023}.

\subsection{Symplectic integration and symmetric projection}

To obtain a solution in the original (non-extended) phase space, one approach involves projecting onto the averaged trajectory once the computation is finished, where the final solution is defined as $(\mathbf{z} + \bar{\mathbf{z}})/2$. Another straightforward alternative is to perform the midpoint projection at each step of the integration~\cite{Luo_2017}. While this \textit{midpoint projection} is computationally efficient and explicit, it does not preserve the symplectic structure in the reduced phase space. We also note that the midpoint projection breaks the time reversibility of its restraint alternative. To recover symplecticity, a \textit{symmetric projection} was introduced in Ref.~\cite{Jayawardana_2023} based on efs.~\cite{Ascher_1999,Hairer_2000,Hairer2006} 
(see also Ref.~\cite{McLachlan_2025}). 

The symmetric projection method determines the solution implicitly at each integration step from $t$ to $t+\tau$. In fact, we seek a correction vector $\bm{\mu}^*$ which is a solution of the consistency condition:
\begin{equation}
    \label{eqn:consistency}
    \mathbf{z}(t+\tau; \mathbf{Z}_0(\bm{\mu}))+ \bm{\mu} = \bar{\mathbf{z}}(t+\tau; \mathbf{Z}_0(\bm{\mu})) -\bm{\mu} ,
\end{equation}
where the initial condition for the extended system is perturbed as $\mathbf{Z}_0(\bm{\mu}) = (\mathbf{z}(t)+\bm{\mu}, \bar{\mathbf{z}}(t)-\bm{\mu})$. The resulting symplectic solution at the new time step $t+h$ is then given by $\mathbf{z}(t+\tau; \mathbf{Z}_0(\bm{\mu}^*)) + \bm{\mu}^*$. Given the consistency condition~\eqref{eqn:consistency}, the two copies at time $t+\tau$, namely $\mathbf{z}(t+\tau; \mathbf{Z}_0(\bm{\mu}^*)) + \bm{\mu}^*$ and $\bar{\mathbf{z}}(t+\tau; \mathbf{Z}_0(\bm{\mu}^*)) - \bm{\mu}^*$, are then equal. The implicit symmetric projection is solved at each step of the integration, using a combination of a fast determination of an initial condition for $\bm{\mu}$ following Refs.~\cite{Hairer_2000,Hairer2006}, and then a Broyden method to refine it with a quadratic convergence. This approach follows the one implemented in Ref.~\cite{Jayawardana_2023}. The main advantage of the symmetric projection is that the integration is symplectic in the original phase space. 

Below, we consider the two projection methods mentioned above in the ${\bf E}\times {\bf B}$ guiding-center model and compare their results with the no-projection case, but with a non-zero restraint coefficient. The two projection methods, midpoint and symmetric, are implemented in the open-source Python package~\cite{pyhamsys} (version 0.87 or higher) for symplectic integration of Hamiltonian systems.  

\section{E\texorpdfstring{$\times$}{x}B model} \label{sec:ExB}

As a first example, we examine the motion of a charged particle in a strong, constant, and uniform magnetic field, and subjected to a turbulent electrostatic potential $V(x,y,t)$. Under the assumptions of a small Larmor radius and high Larmor frequency, guiding center theory allows us to separate the particle’s motion into a rapid gyration around the magnetic field lines and a slower drift perpendicular to them~\cite{Cary2009}. This reduction replaces the original particle by a fictitious one—the guiding center—whose dynamics are governed by the 
${\bf E}\times {\bf B}$ drift. The resulting equation of motion for the guiding center position ${\bf x}=(x,y)$ is:
$$
\dot{\bf x} = \frac{{\bf E}({\bf x},t)\times {\bf B}}{B^2}.
$$
The dynamics of the guiding centers is Hamiltonian, reflecting the Hamiltonian character of the dynamics of the charged particle subjected to the Lorentz force. The Poisson bracket is canonical in the variables $(x,y)$,
and the Hamiltonian is the electrostatic potential $H(x,y,t)=V(x,y,t)$. 

To integrate the equations of motions for the ${\bf E}\times {\bf B}$ model, we extend phase space by considering functions of two copies of the guiding centers,  $(x,y)$ and $(\bar{x},\bar{y})$. The autonomized Hamiltonian is given by Eq.~\eqref{eq:autonomized_extended_Hamiltonian},
with the extended Poisson bracket acting on functions of $\overline{F}(x,y,\bar{x},\bar{y},t)$ given by Eq.~\eqref{eq:autonomized_extended_Poisson_bracket}.
We split the Hamiltonian in three parts: the two potential terms, $V(x,\bar{y},t)$ and $V(\bar{x},y,t)$, and the restraint. 
To test the numerical scheme, we consider a mock potential of the form
$$
    V(x,y,t) = \sum_{\substack{n,m=1 \\ n^2+m^2\leq M^2}} \frac{A}{(n^2+m^2)^{3/2}} \sin (n x+m y +\varphi_{nm}-t),
$$
with $M=25$ and $\varphi_{nm}$ random phases. These random phases are fixed once and for all. For this potential, we investigate the impact of the choice of timestep $dt$ and restraint coefficient $\omega$, and their dependence on the level of local hyperbolicity. We monitor the error in energy as defined by Eq.~\eqref{eq:energy_conservation}. This error in energy is normalized by the number of trajectories. 
We also monitor the distance $\eta(t)={\rm max}_t (\Vert x -\bar{x} \Vert^2 + \Vert y -\bar{y} \Vert^2)^{1/2}$ between the two copies of the guiding center trajectories if no projection is applied. If a symmetric projection is applied, the distance is defined as $\eta(t)={\rm max}_t \Vert {\mu}(t) \Vert$. 

\begin{figure}[htb]
    \centering
    \includegraphics[width=\linewidth]{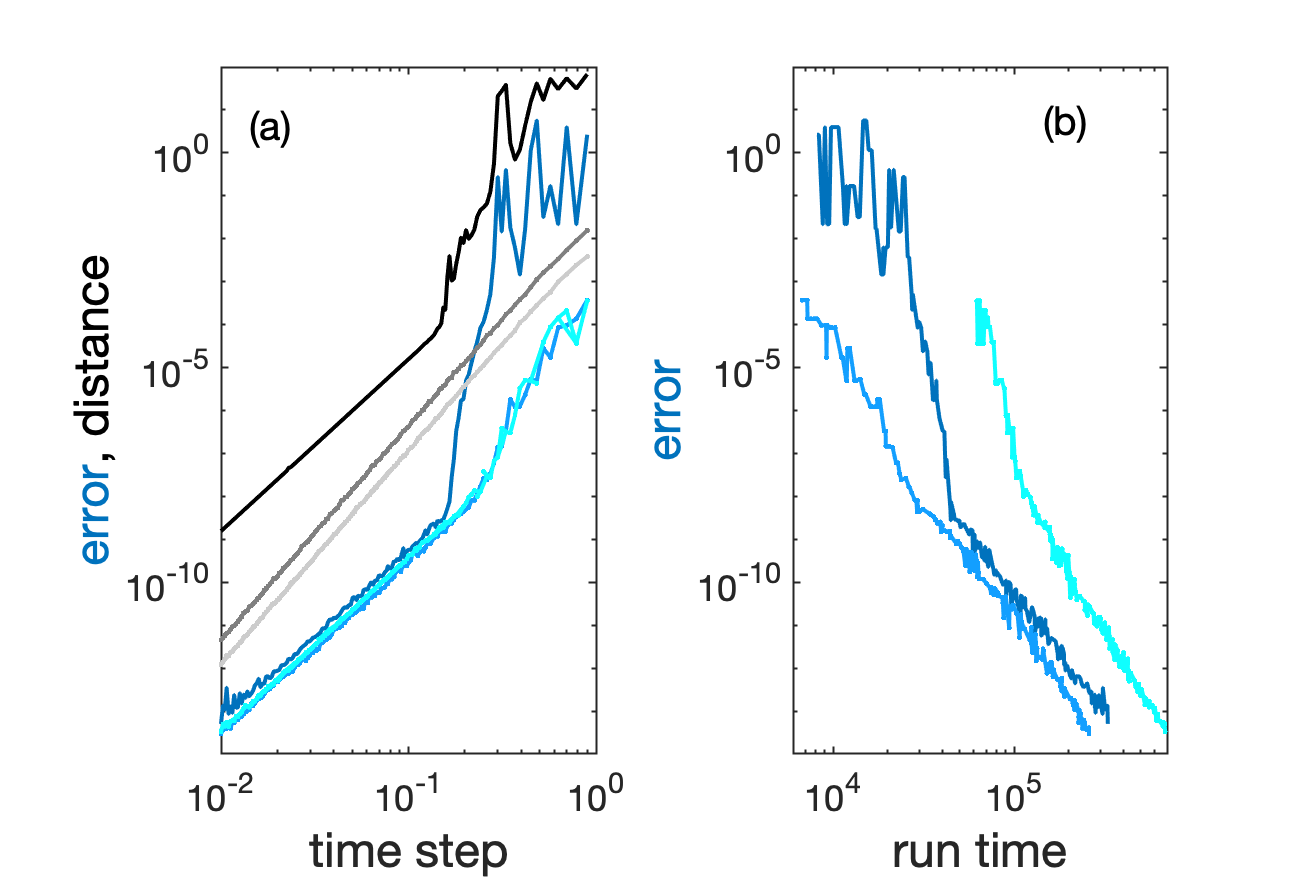}
    \caption{\label{fig:ExB1} 
    Comparison of the error in energy given by Eq.~\eqref{eq:energy_conservation} (blue curves) and averaged distance between copies (black and gray curves) of extended phase-space symplectic integrations as functions of (a) the time step and (b) run time (in seconds). The dark blue and black curves are for the restraint method. The light blue and gray curves are for the midpoint projection method. The cyan and light gray are for the symmetric projection method.
    For all calculations, we use the optimized 4$^\text{th}$ order Blanes and Moan scheme and a restraint coefficient $\omega=10$. The ${\bf E}\times {\bf B}$ model parameters are $A=0.6$ and $M=25$. We use 500 trajectories integrated over 500 periods of the electrostatic field. All quantities are dimensionless.}
\end{figure}

Figure~\ref{fig:ExB1} shows the accuracy and efficacy of extended symplectic integration as functions of (a) the propagation time step and (b) the CPU run time, for an ensemble of 500 trajectories with initial conditions randomly selected over $0\leq x,y\leq2\pi$ and propagated over 500 periods of the electrostatic potential. For each, we compare the results as measured by the phase-space distance between the copies (black and gray curves) and conservation of energy as defined in Eq.~\eqref{eq:energy_conservation} (blue curves).
In panel~(a) we see that, below a time step $dt\approx0.15$, all the error in energy curves exhibit a clear $4^{\rm th}$-order convergence rate matching our choice of split operator scheme. Above $dt\approx0.15$, both the distance and error quickly degrade -- faster than $4^{\rm th}$ power -- and the mapping associated with the propagation scheme loses track of the true ${\bf E}\times {\bf B}$ dynamics. Even though this quick decrease is present for all the methods, we notice that this is much less pronounced for the midpoint and symmetric projection methods. In addition, when the restraint method follows this $4^{\rm th}$-order trend (below a time step of 0.15), all the methods (restraint, midpoint and symmetric projection) provide approximately the same error in energy. However, what distinguishes them is the CPU time necessary to reach that precision in energy. 

Figure~\ref{fig:ExB1}~(b) investigates the efficacy of the schemes, defined as the CPU time required to reach a certain accuracy in the simulation. As expected, the run time necessary to achieve a certain accuracy increases with the required accuracy. One interesting feature for the restraint method is the cross-over between a sharp decrease between time steps $0.3$ and $0.15$ for which the needed CPU time increase is modest (less than double), but the gain in accuracy is significant (more than 8 orders of magnitude). A similar but much less pronounced cross-over is observed for the midpoint and symmetric projection methods in the range of error between $3\times 10^{-4}$ and $2\times 10^{-8}$. This cross-over could help define an optimal trade-off between efficacy and accuracy, which is here found at $dt\approx 0.15$. For comparison, a time step of $0.15$ corresponds approximately to 40 intermediate points in one period of the electrostatic potential. This optimal trade-off is also at $dt\approx 0.15$ for the projection methods.
We notice that the time required for the symmetric projection is about 3 to 5 times higher than the restraint method and the midpoint projection method. For instance, to reach an error in energy of $10^{-10}$, about 55 hours are needed for the symmetric projection, whereas 18 hours are sufficient for the restraint and midpoint projection methods. This additional amount of time is due to the additional implicit equation~\eqref{eqn:consistency} to be solved. 

\begin{figure}[htb]
    \centering
    \includegraphics[width=\linewidth]{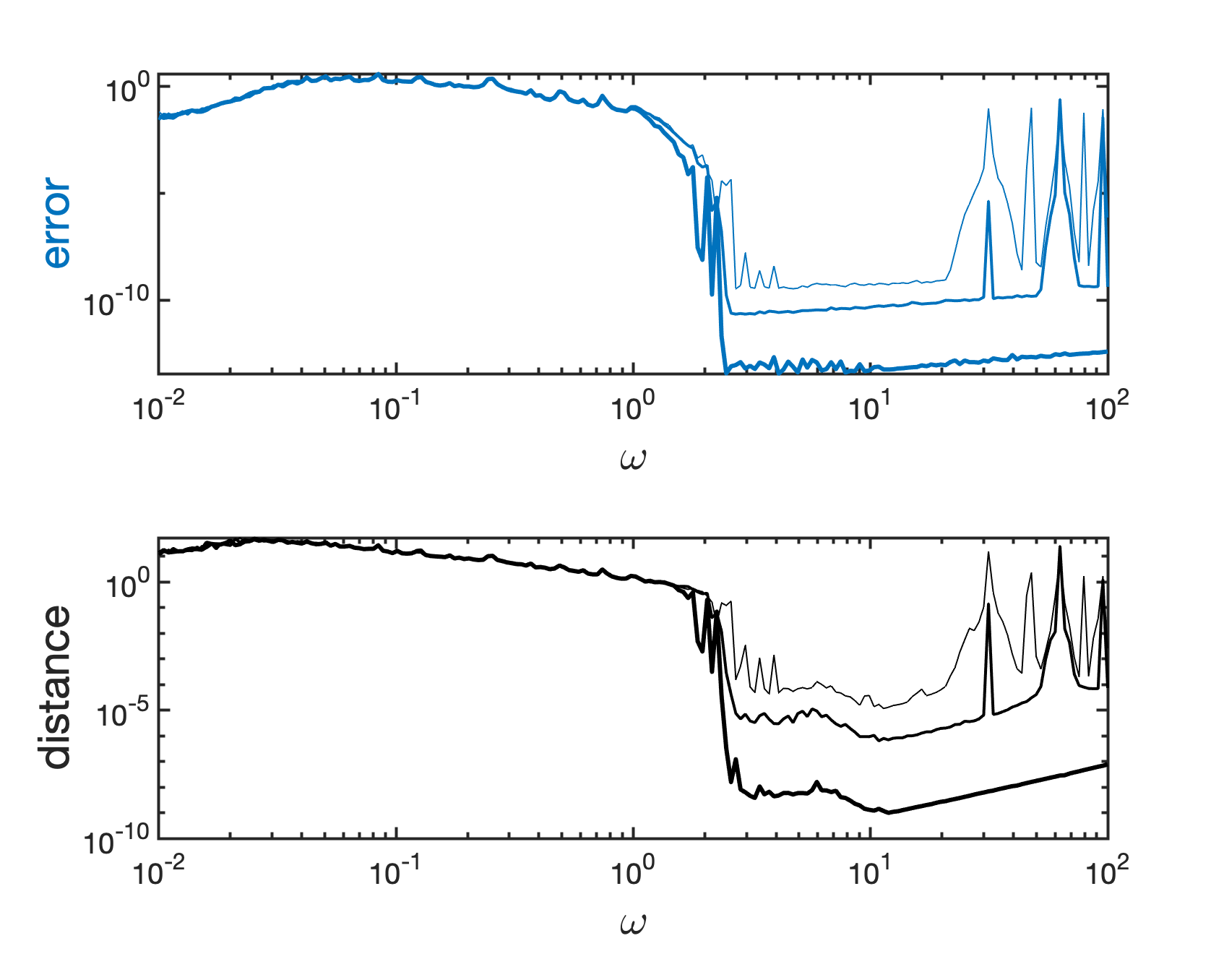}
    \caption{\label{fig:ExB2} 
    Comparison of the error in energy (normalized by the number of trajectories) given by Eq.~\eqref{eq:energy_conservation} (upper panel) and distance between copies (black) of extended phase-space symplectic integrations as functions of $\omega$ for 3 different time steps: $dt=10^{-1}$ (thin line), $dt=5\times10^{-2}$ (regular), and $dt=10^{-2}$ (bold). For information, the error for $dt=10^{-1}$, $dt=5\times10^{-2}$ and $dt=10^{-2}$ is, respectively, $4.3\times10^{-10}$, $2.3\times10^{-11}$ and $3.4\times10^{-14}$ with the symmetric projection, and $2.9\times 10^{-10}$, $1.8\times 10^{-11}$ and $3\times 10^{-14}$ with the midpoint projection.
    We use the same split-operator scheme, model parameters, and trajectory set as in Fig.~\ref{fig:ExB1}.}
\end{figure}

\begin{figure}[htb]
    \centering
    \includegraphics[width=\linewidth]{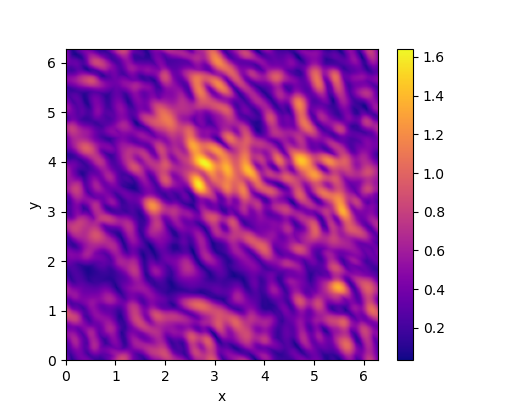}
    \caption{\label{fig:ExB3} Critical restraint coefficient ${\rm max}_t\ \omega_c(x,y,t)$ given by Eq.~\eqref{eq:ExB_wc}, calculated from the local-stability analysis of section~\ref{sec:Methods:restrain_influence}, as a function of $x$ and $y$. We use the same model parameters as in Figs.~\ref{fig:ExB1} and ~\ref{fig:ExB2}.}
\end{figure}

It should be noticed that all these extended phase space method, with or without projection, do not exhibit a drift in energy contrary to conventional integrators implemented directly in the original phase space. 
Next, we investigate the influence of the restraint coefficient~$\omega$ on the stability and accuracy of the integration and show the results in Fig.~\ref{fig:ExB2}. Specifically, for the same set of 500 trajectories as above, we track the conservation of energy given by Eq.~\eqref{eq:energy_conservation} (upper panel) and distance between copies (lower) for three different time steps 0.1, 0.05 and 0.01, while varying the restraint coefficient.
In both panels, unstable integration corresponds to a range of restraint coefficients where the energy error/distance is large compared to their baseline. This figure shows several notable results: First, the shapes of the distance curves track that of the error in energy, which confirms that the distance can be used as a numerically inexpensive gauge for the accuracy of simulations. Second, we see a sharp transition at $\omega\approx2$ between the ranges of restraint coefficient where the integration is stable vs unstable and the transition is essentially independent of the propagation time step, similar to the local stability analysis of section~\ref{sec:Methods:restrain_influence}.
Lastly, above $\omega\approx20$ the energy and distance error degrade again, featuring a succession of sharp peaks for intermediate values of the time steps. We observe fewer peaks with decreasing time steps, which suggest that these are associated with resonances in the discrete map associated with the propagation scheme. By effectively identifying these resonances, the distance measure serves as a diagnostic tool, providing a clear warning when an inappropriate restraint coefficient is being utilized. 
To conclude our analysis of the ${\bf E}\times {\bf B}$ model, we further investigate the relationship between the local stability in the extended phase space analysis of section~\ref{sec:Methods:restrain_influence} with the critical restraint coefficient for which we observe stable integration in simulations -- the sharp drop in the error curves around $\omega\approx2$ in Fig.~\ref{fig:ExB2}. To do this, we compute the map
\begin{eqnarray} \label{eq:ExB_wc}
    &&\omega_{\rm c}(x,y,t) = \frac{1}{4} \left[ -\frac{\partial^2V}{\partial x^2} -\frac{\partial^2V}{\partial y^2}\right. \nonumber \\
    &&\qquad  + \left. \left(\left( \frac{\partial^2 V}{\partial x^2} -\frac{\partial^2 V}{\partial y^2}\right) +4 \left(\frac{\partial^2 V}{\partial x\partial y}\right)^2 \right)^{1/2} \right],
\end{eqnarray}
for which the stability condition given by Eq.~\eqref{eq:example_stability_condition} is satisfied when $\omega>\omega_{\rm c}(x,y,t)$.
Figure~\ref{fig:ExB3} shows the time maximum ${\rm max}_t\ \omega_c(x,y,t)$ as a function of $x$ and $y$. Trajectories in regions of $(x,y)$ with low values of $\omega_c$ (dark purple in the map) do not require a high value of the restraint coefficient $\omega$ to be computed accurately. On the other hand, the local instability regions containing light yellow regions in the map are the ones where the two copies of the trajectories experience stretching, and therefore require a higher value of $\omega$ to damp this divergence. 
The minimum value of the map is approximately 0.03, which is the value of $\omega$ at which the distance between the two copies gradually (linearly) decreases (see lower panel in Fig.~\ref{fig:ExB2}). It takes a higher value of $\omega$ for this decrease to be reflected in the error in energy (see upper panel in Fig.~\ref{fig:ExB2}).  
The ${\bf E}\times {\bf B}$ model exhibits relatively strong global mixing properties, where all trajectories effectively visit the entire $(x,y)$ phase space. In turn, it means that the critical restraint coefficient is essentially trajectory independent, as we have observed in our simulations (not shown), and determined by the overall maximum $\omega_c\approx1.64$ in the map. Notably this global maximum in the map is consistent with the restraint analysis of Fig.~\ref{fig:ExB2} and closely match the value of $\omega\approx 2$ discussed above. 
Given the equation for the critical restraint coefficient $\omega_c$ given by Eq.~\eqref{eq:ExB_wc}, we expect it to depend linearly on $A$, \textit{i.e.}, a higher value of $A$ will require a higher value of the restraint coefficient $\omega$ to reach the same accuracy.  
We expect that extending the local-stability analysis to systems with segregated regions of phase space—such as those exhibiting mixed elliptic and hyperbolic dynamics, or multiple chaotic domains separated by transport barriers—will confirm that the stability condition within each region is determined by its corresponding local restraint coefficient $\omega_c$.

\section{TDDFT dynamics} \label{sec:TDDFT}

In this section, we explore the application of the extended phase-space technique for simulation of TDDFT dynamics. 
For clarity, we first review and discuss the restraint case and then compare its performance with the midpoint projection in section~\ref{sec:TDDFT:midpoint}. We note that the latter formally corresponds to taking $\omega=0$ and performing the projection at the end of each propagation time step, such that the methodology development discussed in section~\ref{sec:TDDFT:extended} broadly apply to both methods. Given its significant increased numerical complexity and run time, we did not implement the symmetric projection for TDDFT.
To streamline discussions, we often provide subset of equations that are relevant to illustrate the points being discussed rather than the entire sets of equations for the dynamics. In appendix~\ref{app:TDDFT_split_operator_equations} we provide a comprehensive summary of the equations that must be implemented for the extended phase-space split operator schemes we put forward in this section.
We have implemented all the TDDFT split-operator propagators we discuss in this Paper in version~1.24 of the \texttt{QMol-grid} package~\cite{mauger_2024}, which is available for download on its GitHub and MathWorks repositories~\cite{QMol_grid}.
To speed up calculations, our implementation takes advantage of the refactorization discussed around Eq.~\eqref{eq:symplectic_split_operator}.

\subsection{Hamiltonian structure of TDDFT} \label{sec:TDDFT:Hamiltonian_structure}

We laid out the Hamiltonian structure of Kohn-Sham (KS) TDDFT in detail in Ref.~\cite{Mauger_2023} and only summarize here the important elements for understanding the application of the extended phase space method. 
KS TDDFT is described by the nonlinear systems of coupled partial differential equations
\begin{equation} \label{eq:TDDFT}
    i\partial_{t} \phi_k \left({\bf x};t\right) = 
        \hat{\mathcal{H}}_\text{eff}\left[\left\{\phi_k\right\}_k\right] \phi_k\left({\bf x};t\right),
\end{equation}
where ${\bf x}\in {\mathbb R}^d$ and $\hat{\mathcal{H}}_\text{eff}$ is the DFT Hamiltonian operator, which explicitly depends on the field variables $\left\{\phi_k\right\}_k$. 
The $\phi_k$ are called the KS orbitals and describe the quantum electronic structure/dynamics via the one-body density -- see Eq.~\eqref{eq:one-body_density} below.
For simplicity, here we omit spin in the KS orbitals that would otherwise leads to more cumbersome yet similar expressions in the equations below. Likewise, we consider field-free TDDFT dynamics while adding the action of an external driving electric field in the dipole approximation is straightforward.
Physical $\hat{\mathcal{H}}_\text{eff}$ in Eq.~\eqref{eq:TDDFT} are Hermitian operators and the KS orbitals form a time-dependent family of orthonormal functions
\begin{equation} \label{eq:KS_are_orthonormal}
    \int{\phi_k\left({\bf x};t\right)^* \phi_l\left({\bf x};t\right) {\rm d}{\bf x}} = \delta_{k,l} \qquad \forall k,l \textrm{ and } \forall t.
\end{equation}
The DFT Hamiltonian operator is usually decomposed into its kinetic, external, Hartree, and exchange-correlation components as
\begin{equation} \label{eq:TDDFT_Hamiltonian_decomposition}
    \hat{\mathcal{H}}_\text{eff} = 
        -\frac{\Delta}{2} + \hat{\mathcal{V}}_{\rm ext} +
        \hat{\mathcal{V}}_{H} + 
        \hat{\mathcal{V}}_{\rm XC}.
\end{equation}
The external potential operator $\hat{\mathcal{V}}_{\rm ext}$ may include an explicit time dependence, {\it e.g.}, modeling the interaction with an external driving electric field. 
Next, the Hartree potential operator $\hat{\mathcal{V}}_{\rm H}$ corresponds to the classical electrostatic interaction of each KS orbital with the mean-field density generated by all the electrons in the model. 
Finally, the exchange-correlation potential operator $\hat{\mathcal{V}}_{\rm XC}$ collects all the non-classical electron-electron interactions.

Next, we introduce the Hamiltonian structure of KS TDDFT. To simplify notations, in what follows we omit the implicit time dependence of the KS orbitals  $\phi_k({\bf x})\equiv\phi_k({\bf x};t)$ and all other dynamical variables. 
Briefly, in the Hamiltonian formulation of TDDFT the Poisson algebra is defined as the set $\mathcal{A}$ of the functionals of the field variables $\phi_k({\bf x})$ and $\phi_k^*({\bf x})$.
The Hamiltonian functional $H\left[\{\phi_k\},\{\phi_k^*\}\right]$ corresponds to the standard DFT total-energy functional with the one-body density functionals  defined as 
\begin{equation} \label{eq:one-body_density}
    \rho = \sum_{k}{n_k\ \phi_k^* \phi_k},
\end{equation}
where $n_k$ is the population (number of electrons) associated with the $k^\text{th}$ KS orbital and involving the pairs of canonically conjugate field variables $\{\phi_k\}_k$ and $\{\phi^*_k\}_k$. 
For example, the kinetic and external-potential parts of the TDDFT Hamiltonian functional are respectively given by
$$
    H_\text{kin} = -\frac{1}{2} \sum_k{n_k \int{\phi_k^* \Delta \phi_k}}~{\rm d}{\bf x} \quad \text{and} \quad
    H_\text{ext} = \int{\mathcal{V}_\text{ext} \rho}~{\rm d}{\bf x}.
$$
For the other components, the functionals are obtained by substituting density components $|\phi_k|^2$ with the product of their associated conjugate phase-space variables $\phi_k\phi_k^*$ in conventional DFT energy functionals~\cite{Mauger_2023}, owing that physical systems enforce the symmetry $\phi_k=\left(\phi_k^*\right)^\dagger$ with ``$^\dagger$'' denoting the complex conjugate.

The TDDFT canonical Poisson bracket is
\begin{equation} \label{eq:TDDFT_Poisson_bracket}
    \left\{F,G\right\} = \sum_k{ \frac{1}{n_k} \int{ \frac{1}{i}  \left(
        \frac{\delta F}{\delta \phi_k^*} \frac{\delta G}{\delta \phi_k} - \frac{\delta F}{\delta \phi_k} \frac{\delta G}{\delta \phi_k^*}
        \right) {\rm d} {\bf x}}},
\end{equation}
for any two functionals $F$ and $G$.
For virtual KS orbitals where $n_k=0$ we take the convention $n_k/n_k=1$, where the population coefficient in the Poisson bracket~\eqref{eq:TDDFT_Poisson_bracket} cancels with that of the functional derivative, to match the limit $n_k\to0$.
One can easily check that the Poisson bracket is antisymmetric, bilinear,
and satisfies Leibniz rule and Jacobi identity.
We recover the KS TDDFT equations~\eqref{eq:TDDFT} using the generic Hamiltonian flow formula~\eqref{eq:Hamiltonian_flow} with $F=\phi_k$ and $F=\phi_k^*$
\begin{eqnarray*}
    && \dot{\phi}_k = \{H, \phi_k\} = 
        +\frac{1}{i n_k} \frac{\delta H}{\delta \phi_k^*} = 
        -i \hat{\mathcal{H}}_\text{eff} \left[\{\phi_k\},\{\phi_k^*\}\right] \phi_k, \\
    && \dot{\phi}_k^* = \{H, \phi_k^*\} = 
        -\frac{1}{i n_k} \frac{\delta H}{\delta \phi_k} = 
        + i \hat{\mathcal{H}}_\text{eff} \left[\{\phi_k\},\{\phi_k^*\}\right] \phi_k^*,
\end{eqnarray*}
using the fact that physical Hamiltonian operators are Hermitian $\hat{\mathcal{H}}_\text{eff}^\dagger=\hat{\mathcal{H}}_\text{eff}$.
One may also break down the Hamiltonian functional $H$ between the kinetic and various potential functionals, as in Eq.~\eqref{eq:TDDFT_Hamiltonian_decomposition}, and recover their associated operator components
\begin{eqnarray*}
    && \{H_\text{el}, \phi_k\} = 
        +\frac{1}{i n_k} \frac{\delta H_\text{el}}{\delta \phi_k^*} = 
        -i \hat{\mathcal{H}}_\text{el} \left[\{\phi_k\},\{\phi_k^*\}\right] \phi_k, \\
    && \{H_\text{el}, \phi_k^*\} = 
        -\frac{1}{i n_k} \frac{\delta H_\text{el}}{\delta \phi_k} = 
        + i \hat{\mathcal{H}}_\text{el} \left[\{\phi_k\},\{\phi_k^*\}\right] \phi_k^*,
\end{eqnarray*}
for all $k$, where $H_\text{el}$ is any of $H_\text{kin}$, $H_\text{ext}$, $H_\text{H}$, or $H_\text{XC}$ functionals, respectively generating the kinetic, external, Hartree, and exchange-correlation operators.

\subsection{Extended TDDFT phase space} \label{sec:TDDFT:extended}

Following the procedure of section~\ref{sec:Methods:Extended_phase_space}, we extend phase space to the pairs of KS orbital variables $\phi_k$, $\overline{\phi}_k$, $\phi_k^*$, and $\overline{\phi}_k^*$ together with the extended Poisson bracket
\begin{eqnarray}
    &&\left\{F,G\right\}  = \sum_k{ \frac{1}{n_k} \int{ \frac{1}{i}  \left(
        \frac{\delta F}{\delta \phi_k^*} \frac{\delta G}{\delta \phi_k} - \frac{\delta F}{\delta \phi_k} \frac{\delta G}{\delta \phi_k^*}
        \right) {\rm d} {\bf x}}} \nonumber \\ &&
        \qquad + \sum_k{ \frac{1}{n_k} \int{ \frac{1}{i}  \left(
        \frac{\delta F}{\delta \overline{\phi}_k^*} \frac{\delta G}{\delta \overline{\phi}_k} - \frac{\delta F}{\delta \overline{\phi}_k} \frac{\delta G}{\delta \overline{\phi}_k^*}
        \right) {\rm d} {\bf x}}}. \label{eq:TDDFT_extended_Poisson_bracket}
\end{eqnarray}
For the grid-based discretization and fast-Fourier transform calculation of differential operators we use in our TDDFT code~\cite{mauger_2024}, we have found that the Hamiltonian extension $\overline{H}_a[\phi,\phi^*,\overline{\phi},\overline{\phi}^*]=H[\phi,\overline{\phi}^*]+H[\overline{\phi},\phi^*]+R[\phi,\phi^*,\overline{\phi},\overline{\phi}^*]$ is unconditionally unstable. 
Specifically, the instability comes from the kinetic component of the Hamiltonian where, for instance,
$$
    \text{e}^{\tau \mathcal{L}_{H_\text{kin}[\phi,\overline{\phi}^*]}} \overline{\phi}_k = 
        \overline{\phi}_k + i \tau \frac{\Delta}{2} \phi_k = 
        \overline{\phi}_k - i \tau \mathcal{F}^{-1}\left[\frac{{\bf p}^2}{2}\mathcal{F}[\phi_k]\right],
$$
where $\mathcal{F}$ denotes the Fourier transform and we see that high spatial frequencies ${\bf p}$ are amplified by the (linearized) exponential.
Alternatively, keeping the pairs of conjugate KS orbitals together in the kinetic terms $H_\text{kin}[\phi,\phi^*]$ and $H_\text{kin}[\overline{\phi},\overline{\phi}^*]$  yields
$$
    \text{e}^{\tau \mathcal{L}_{H_\text{kin}[\phi,\phi^*]}} \overline{\phi}_k = 
        \text{e}^{i \tau \frac{\Delta}{2}} \phi_k = 
        \mathcal{F}^{-1}\left[\text{e}^{-i \tau \frac{{\bf p}^2}{2}}\mathcal{F}[\phi_k]\right],
$$
which is unitary (and likewise for the other extended phase-space variables).
All in all, the stabilized extended phase-space TDDFT Hamiltonian reads
\begin{eqnarray}
    \overline{H}_a[\phi, \phi^*, \overline{\phi}, \overline{\phi}^*] & = & 
        H_\text{kin}[\phi,\phi^*] + H_\text{kin}[\overline{\phi},\overline{\phi}^*] \nonumber \\ &&
         + H_\text{pot}[\phi,\overline{\phi}^*] + H_\text{pot}[\overline{\phi},\phi^*] \nonumber \\ &&
        + R[\phi, \phi^*, \overline{\phi}, \overline{\phi}^*], \label{eq:TDDFT_extended_Hamiltonian}
\end{eqnarray}
where $H_\text{pot}$ is the functional associated with the total KS potential, {\it i.e.}, the external, Hartree and exchange-correlation potentials in Eq.~\eqref{eq:TDDFT_Hamiltonian_decomposition}.

The last missing piece in the extended phase-space Hamiltonian functional~\eqref{eq:TDDFT_extended_Hamiltonian} is the restraint $R$, which we take as
\begin{equation} \label{eq:TDDFT_restrain}
    R[\phi,\phi^*,\overline{\phi},\overline{\phi}^*] = \omega \sum_k{n_k \int{
            (\phi_k-\overline{\phi}_k)(\phi_k^*-\overline{\phi}_k^*)\ {\rm d}{\bf x}
        }}.
\end{equation}
Like in the other pieces of the Hamiltonian functional, the populations coefficients $n_k$ in the restraint ensure the cancellation of the $1/n_k$ from the Poisson bracket~\eqref{eq:TDDFT_extended_Poisson_bracket}. Physically, it ensures that virtual orbitals that do not carry any electrons and thus do not participate in the dynamics are not artificially influencing the restraint.
Similar to Eq.~\eqref{eq:extended_restrain_flow}, the flow generated by the TDDFT restraint can be integrated analytically and reads
\begin{subequations} \label{eq:TDDFT_restrain_flow}
\begin{align}
    & \text{e}^{\tau {\mathcal{L}}_{R}}\phi_k = 
        \frac{\phi_k+\overline{\phi}_k}{2} + 
        \text{e}^{-2 i \omega \tau} \frac{\phi_k-\overline{\phi}_k}{2}, \\
    & \text{e}^{\tau {\mathcal{L}}_{R}}\phi_k^* = 
        \frac{\phi_k^*+\overline{\phi}_k^*}{2} + 
        \text{e}^{-2 i \omega \tau} \frac{\phi_k^*-\overline{\phi}_k^*}{2},
\end{align}
\end{subequations}
and changing the sign in from of the complex exponential for $\overline{\phi}$ and $\overline{\phi}^*$.
For $\omega=0$ the pair of equations simplify to the identify, reflecting the fact that no restraint is applied.

To conclude this section, we note that the KS potential can formally be decomposed between an ``explicit'' part, that is a functional of the one-body density alone, and and ``implicit'' part that depends on the KS orbitals 
$$
    H_\text{pot}[\phi,\phi^*] = H_\text{exp}[\rho] + H_\text{imp}[\phi,\phi^*],
$$
with the one-body density functional $\rho$ defined in Eq.~\eqref{eq:one-body_density}.
In Ref.~\cite{Mauger_2023} we have shown that, for grid-based discretization schemes, the flow generated by the explicit part is analytically solvable and reads
$$
    \text{e}^{\tau\mathcal{L}_{H_\text{exp}[\rho]}} \phi_k = 
        \text{e}^{-i\tau V_{exp}[\rho]} \phi_k, \quad \text{with} \quad
        V_\text{exp}[\rho] = \frac{\delta H_\text{exp}}{\delta \rho}.
$$
This suggests that we may treat the explicit part of the potential like the kinetic term above, keeping the pairs of conjugate variables together, and only treat the implicit part with the mixed extend phase-space variables. The corresponding extended phase-space Hamiltonian then becomes
\begin{eqnarray}
    \overline{H}_a[\phi, \phi^*, \overline{\phi}, \overline{\phi}^*] & = & 
        H_\text{kin}[\phi,\phi^*] + H_\text{kin}[\overline{\phi},\overline{\phi}^*] \nonumber \\ &&
        + H_\text{exp}[\rho(\phi,{\phi}^*)] + H_\text{exp}[\rho(\overline{\phi},\overline{\phi}^*)] \nonumber \\ &&
        + H_\text{imp}[\phi,\overline{\phi}^*] + H_\text{imp}[\overline{\phi},\phi^*] \nonumber \\ &&
        + R[\phi, \phi^*, \overline{\phi}, \overline{\phi}^*], \label{eq:TDDFT_extended_Hamiltonian_split_potential}
\end{eqnarray}
In what follows, we refer to this version of the extended Hamiltonian as ``split potential'' (or ``split V'' for short) and to the one of Eq.~\eqref{eq:TDDFT_extended_Hamiltonian} when we do not mention any split.

\subsection{Numerical simulations} \label{sec:TDDFT:simulation}

In the previous section, we laid out the solutions for the flow generated by the different components in the extended TDDFT Hamiltonian~\eqref{eq:TDDFT_extended_Hamiltonian} and~\eqref{eq:TDDFT_extended_Hamiltonian_split_potential}.
In principle, we could therefore apply the split-operator methods discussed around Eq.~\eqref{eq:symplectic_split_operator} to solve numerically for the dynamics in the extended phase space and, from there, obtain solutions to the TDDFT equation of motion~\eqref{eq:TDDFT} by averaging, as discussed at the end of section~\ref{sec:Methods:Extended_phase_space}.
In practice, this approach raises difficulties for evaluating some of the terms involved in the split-operator components and requires the calculation of four field variables for each KS orbital in the final averaged solution. In this section we discuss how to avoid these difficulties and with it reduce the computational cost of the propagation.

Because the pairs of extended phase-space conjugate variables are grouped differently in the kinetic and potential parts of the extended Hamiltonian, split-operator schemes have no reason to preserve the complex conjugation $\phi^*\neq \phi^\dagger$ or $\overline{\phi}^*\neq \overline{\phi}^\dagger$.
In turn, it means that the one-body densities $\rho(\phi,\phi^*)$ and $\rho(\overline{\phi},\overline{\phi}^*)$ given by Eq.~\eqref{eq:one-body_density} can have some (residual) imaginary part. Yet, DFT potential functionals are designed with the implicit assumption that the one-body density is real valued, as would be the case for physical systems.
To resolve this conundrum, in simulations we adopt the following procedure: (i) we derive the equations of motion and analytical solutions for the flow generated by the various components of the split-operator scheme using the four independent sets of field variables $\phi$, $\phi^*$, $\overline{\phi}$, and $\overline{\phi}^*$, as detailed in the previous section. Then, (ii) in numerical simulations we systematically replace $\phi^*$ with $\phi^\dagger$ and $\overline{\phi}^*$ with $\overline{\phi}^\dagger$ in the flow solution. For instance
$
    V_\text{exp}\left[\rho(\overline{\phi},\overline{\phi}^*)\right] \to 
        V_\text{exp}[\rho\left(\overline{\phi},\overline{\phi}^\dagger)\right]
$
and
$
    V_\text{exp}\left[\rho(\phi,\phi^*)\right] \to 
        V_\text{exp}[\rho\left(\phi,\phi^\dagger)\right].
$
In turn, it means that we need to propagate only $\phi$ and $\overline{\phi}$ as their conjugate counterpart become irrelevant for the propagation.
As a result, the memory footprint for KS orbitals in the extended split-operator scheme is three times that of the orbital solution: one for each of $\phi$ and $\overline{\phi}$ plus an extra temporary field variable when calculating the restraint or the phase-space average when the orbitals for the non-extended problem are required. For the midpoint projection, one should be able to further reduce the memory footprint to twice the orbitals.

\begin{figure}[htb]
    \centering
    \includegraphics[width=\linewidth]{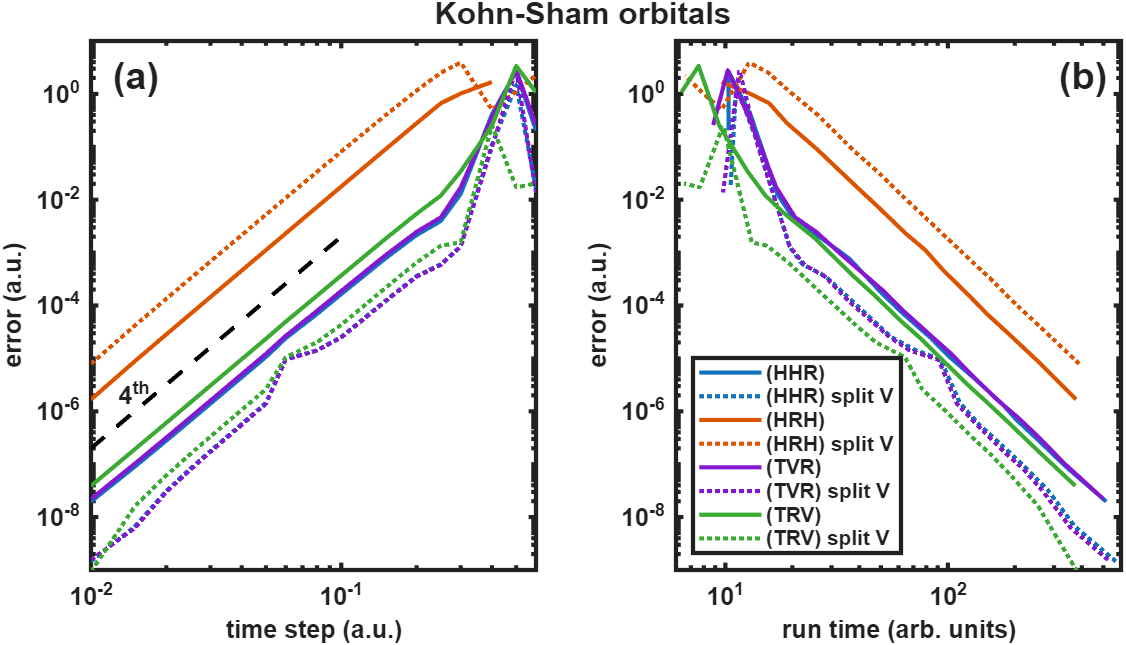}
    \caption{\label{fig:TDDFT_compare_splits} 
    Comparison of the (a) accuracy and (b) efficacy of extended phase-space symplectic integrations using different types and orders of terms in the split operator -- see legend and main text. 
    The HHR and TVR curves are nearly on top of each other.
    For all calculations, we use the optimized 4$^\text{th}$ order Blanes and Moan scheme and a restraint coefficient $\omega=10$. 
    For reference, the dashed black line in panel (a) mark a perfect 4$^\text{th}$ order scaling.}
\end{figure}

Similar to the ${\bf E}\times{\bf B}$ analysis of Fig.~\ref{fig:ExB1}, Fig.~\ref{fig:TDDFT_compare_splits} compares the (a) accuracy and (b) efficacy of far-from-equilibrium TDDFT simulations using the optimized 4$^\text{th}$-order Blanes and Moan split operator~\cite{Blanes_2002} -- see appendix~\ref{app:TDDFT_test_system} for details about the system we use in our tests and how we calculate the error.
When implementing the split operator, we have many possible choice as to the order in which to take the Hamiltonian components of Eq.~\eqref{eq:TDDFT_extended_Hamiltonian}. The order of the operator matters for the accuracy of the computation as it was observed in Ref.~\cite{Fasso_2003}. In Fig.~\ref{fig:TDDFT_compare_splits} we systematically compare four such choices where the letters in the legend represent the order in which the (H) Hamiltonian, (T) kinetic, (V) potential, and (R) restraint are taken in $\chi_\tau$ given by Eq.~\eqref{eq:symplectic_chi_fb}:
\begin{itemize}
    \item HHR: in order, 
        (i)   apply the kinetic and then potential components that update $\phi$, 
        (ii)  apply the kinetic and then potential components that update $\overline{\phi}$, and
        (iii) apply the restraint.
    \item HRH: same as HHR with steps (ii) and (iii) swapped.
    \item TVR: in order
        (i)   apply the kinetic components that update $\phi$ and then $\overline{\phi}$,
        (ii)  apply the potential components that update $\phi$ and the $\overline{\phi}$, and
        (iii) apply the restraint.
    \item TRV: same as TVR with steps (ii) and (iii) swapped.
\end{itemize}
When splitting the potential (dotted curves), we use the separation of the potential between its explicit and implicit components as in Eq.~\eqref{eq:TDDFT_extended_Hamiltonian_split_potential} while keeping the order discussed above.

Figure~\ref{fig:TDDFT_compare_splits}~(a) shows the error in the KS orbitals at the end of the calculation as a function of the propagation time step for the different choices of splitting discussed above.
All curves show a clear 4$^\text{th}$-order convergence, as can be seen by comparison with the dashed black guide line. When using different split-operator schemes, we have systematically observed that the convergence rate matches the scheme order, thus demonstrating the applicability of the propagation methods we put forward in this Paper -- see appendix~\ref{app:split_scheme}.
Compared to the other splits, we see that HRH under-performs by over an order of magnitude. We have observed similar results across all the split-operator schemes we tried. 
For the other split patterns, we see that  splitting the explicit and implicit parts of the potential improves the accuracy by about one order of magnitude -- compare the solid and dotted lines.
Finally, the HHR and TVR patterns have about the same accuracy (the curves are nearly on top of each other on the plot) and are about twice more accurate than TRV.
When exploring different tests, we have found that the relative accuracy of the HHR, TVR, and TRV can change but are systematically close to each other, and that splitting the explicit and implicit parts of the potential yields a significant improvement in the accuracy.

Figure~\ref{fig:TDDFT_compare_splits}~(b) compares the efficacy of the different split-operator flavors, defined as the simulation walltime required to reach a given accuracy. As such, increasing the efficiency corresponds to a leftward shift in the plot.
Unsurprisingly, we see that HRH is the least efficient, as a result of its poorer accuracy compared to the other split motifs. Likewise, we find that splitting the potential between its explicit and implicit components lead to enhanced efficiency for the results -- compare the solid and dotted curves for HHR, TVR, and TRV results. The potential split leads to an increased number of terms in the split operator, and thus increased computations per time steps that are nevertheless largely offset by the increased accuracy. 
Like for the accuracy results, we observe little difference between the efficacies of the HHR and TVR split motifs, with the respective curves being essentially on top of each other in the plot. On the other hand, we see that the TRV split slightly outperforms in terms of efficacy. We attribute this to the refactorization of the potential (resp.\ implicit part of the potential when splitting it) between the $\chi_{\alpha_{2k}\tau}$ and $\chi_{\alpha_{(2k-1)}\tau}^*$ steps of Eq.~\eqref{eq:symplectic_split_operator}. In our test simulations, the (implicit part of the) potential is by far the most expensive component to calculate and thus explain how the refactorization pull TRV as the most efficient. 
We expect the balance between accuracy and refactorization-induced efficiency to be dependent of the choice of DFT functional used in the simulations.

\begin{figure}[htb]
    \centering
    \includegraphics[width=\linewidth]{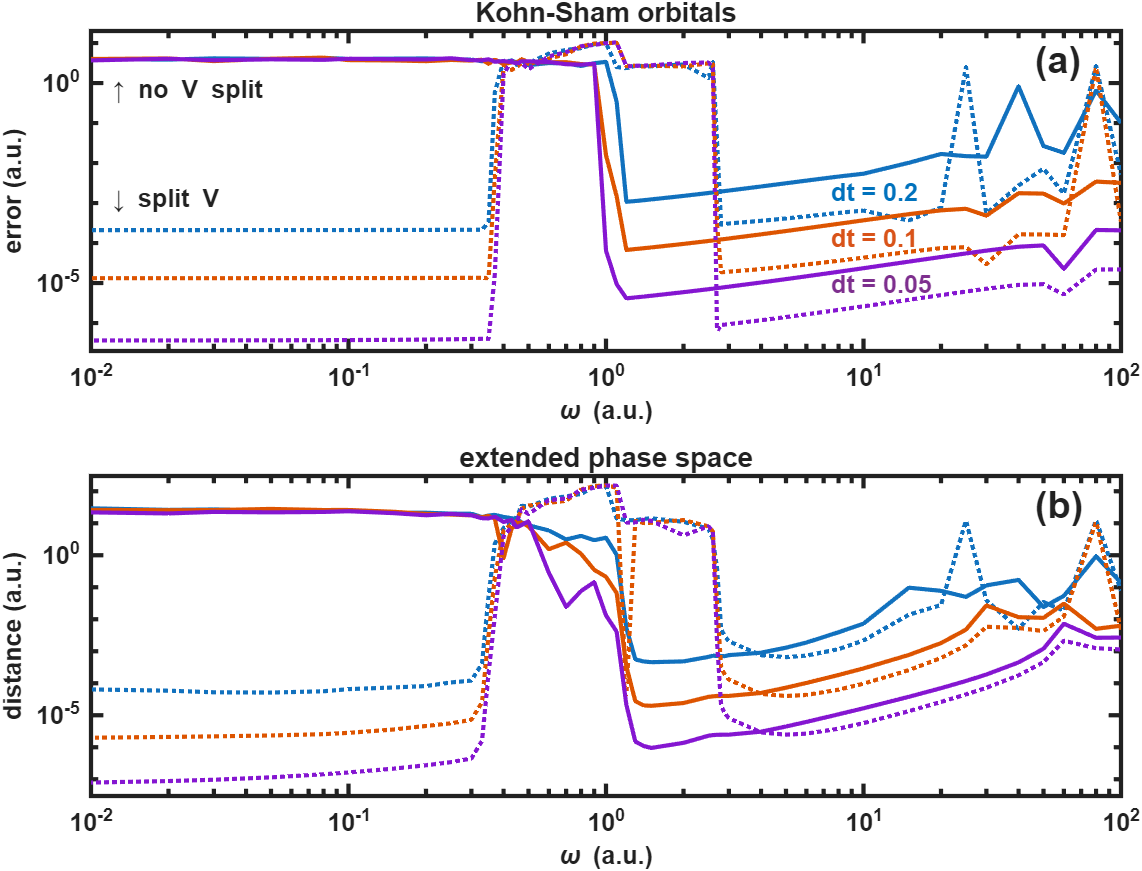}
    \caption{\label{fig:TDDFT_compare_restrains} 
    Comparison of the (a) accuracy and (b) distance between the extended phase-space Kohn-Sham orbitals~\eqref{eq:TDDFT_extended_phase_space_distance} while varying the restraint coefficient $\omega$ for different propagation time steps -- see labels on the curves.
    For all calculations, we use the optimized 4$^\text{th}$ order Blanes and Moan scheme and TRV operator split.}
\end{figure}

Next, we investigate the influence of the restraint coefficient $\omega$ in Eq.~\eqref{eq:TDDFT_restrain} on the stability and precision of the integration and show the results in Fig.~\ref{fig:TDDFT_compare_restrains}. Specifically, for three propagation time steps, we systematically track the (a) error in the final KS orbitals and (b) $L^2$-norm distance between the extended phase-space KS orbitals 
\begin{equation} \label{eq:TDDFT_extended_phase_space_distance}
    d = \sqrt{\sum_k{\int{|\phi_k-\overline{\phi}_k|^2}\ {\rm d}{\bf x}}}    
\end{equation}
as functions of the restraint coefficient. 
In both panels, unstable integration corresponds to range of restraint coefficients where the error/distance is large compared to their baseline. Figure~\ref{fig:TDDFT_compare_restrains} shows several notable results, which are strikingly reminiscent of what we observed for the ${\bf E}\times{\bf B}$ model -- comparing it with Fig.~\ref{fig:ExB2}:
First, the shapes of the distance curves in panel (b) track that of the error in (a), which once again confirms that the distance can be used as a numerically inexpensive gauge for the accuracy of simulations.
Second, we see sharp transitions between the ranges of restraint coefficients where the integration is stable vs unstable and the transitions are essentially independent of the propagation time step. More generally, we have found that these transitions are essentially independent of the split-operator scheme (2$^\text{nd}$-order Strang/Verlet~\cite{Strang_1968}, 4$^\text{th}$-order Forest-Ruth~\cite{Forest_1990}, and optimized 4/6$^\text{th}$-order Blanes-Moan~\cite{Blanes_2002}), as well as the split order (HHR, HRH, TVR, and TRV). These indicate that the stability of the integration is determined by the local stability of the extended phase-space flow, as discussed in section~\ref{sec:Methods:restrain_influence}, rather than the specific scheme and time step used.
Third, we observe different regions of stability depending on whether we split the explicit and implicit parts of the potential (dotted curves) or not (solid). For the latter, we see that $\omega_c\approx 1$ is the threshold above which the propagation is stable and corresponds to case (ii) in the discussion of the influence of the restraint on the dynamic of section~\ref{sec:Methods:restrain_influence}, below Eq.~\eqref{eq:example_stability_condition}. For the former, we observe a region of instability $0.3\lesssim\omega\lesssim3$, which corresponds to case (iii) in the discussion. 
Finally, like in the ${\bf E}\times{\bf B}$ model, we likely attribute the sharp peaks in the error/distance for larger values of $\omega$ to resonances in the discrete map associated with the propagation scheme. Again we note how these peaks are reduced when decreasing the time step.

\subsection{Midpoint projection} \label{sec:TDDFT:midpoint}

Like in the ${\bf E}\times {\bf B}$ example of section~\ref{sec:ExB}, we have systematically compared the accuracy of the extended phase space integration when using the restraint vs midpoint projection and, overall, observe similar results  -- see also appendix~\ref{app:restraint_vs_MPP} and its figure~\ref{fig:TDDFT_restraint_vs_MPP}:
(i) The midpoint projection generally avoids the resonance peaks observed in the restraint case with larger propagation time steps.
(ii) The distance between $\phi_k$ and $\overline{\phi}_k$, right before applying the midpoint projection, tracks the error in the propagation. That distance is a couple of orders of magnitude smaller than in the restraint case, due to the fact that $\phi_k$ and $\overline{\phi}_k$ are reset to matching values at the end of each time step where the restraint let them evolve apart from each other.
(iii) The most accurate restraint case, when splitting the potential and with $\omega=0$, yields essentially identical results to the midpoint projection.

In our TDDFT calculations, one notable difference between the restraint and midpoint projection is the relative insensitivity of the latter to the order of the terms (kinetic, potential, etc.) in the split operator, whereas the restraint yields much poorer performance in the ``HRH'' split -- see figure~\ref{fig:TDDFT_compare_splits} and associated discussion in the main text. This, combined with point (i) above, suggest that the midpoint projection offers a good/safer choice for general-purpose TDDFT calculations. On the other hand, the added $\omega$ parameter in the restraint may offer, in some specific case, an opportunity to optimize the performance of TDDFT calculations.

\section{Summary, conclusions, and outlook} \label{sec:Conclusion}

In summary, we have explored the use of phase-space extension~\cite{Pihajoki2015,Tao_2016} to enable symplectic split-operator integration of Hamiltonian systems with canonical and more generally, constant Poisson brackets, that are otherwise not amendable to conventional split operators schemes.
Using the examples of a one-and-a-half degree of freedom system originating from plasma physics and infinite dimensional model from physical chemistry, we illustrate the generality of the approach and stability condition for numerical integration of the dynamics using high-order symplectic split-operator schemes. Remarkably, despite their profound differences—one being a low-dimensional classical system and the other a high-dimensional quantum mechanical model—both exhibit strikingly similar numerical behaviors, particularly in terms of stability and accuracy as the parameters of the integration schemes are varied.
In the restraint case, following Ref.~\cite{Tao_2016}, we have shown that the restraint coefficient—a key parameter in the scheme—must be carefully selected to ensure local stability in the extended phase space. We have compared the restraint method with a method combining an extension in phase space and a projection, be it a midpoint or implicit symmetric projection. We observed that all these extended phase-space methods are qualitatively similar, providing errors in energy of the same order, and no drift in energy compared to conventional integrators implemented without extension. When an apt projection is used, symmetric or midpoint, we have shown that the methods provide similar results in a broad range of parameters with some variations: The midpoint projection seems very efficient, even though it does not ensure a symplectic (or even energy conserving) integration in the extended phase space and on the original phase space. On the other hand, the symmetric projection ensures the symplectic integration in the original phase space. However, it is computationally expensive given the implicit consistency equation to be solved at each time step, and relies on the quality of the initial guess to converge towards a solution. A compromise is afforded by the use of a restraint, which ensures that the split-integration is symplectic in the extended phase space. However, the integration is not symplectic in the original phase space, and given the introduction of an additional frequency in the system, might exhibit resonances. These large errors in integration are mitigated by the monitoring of a computationally efficient diagnostic, the distance between the two copies in the extended phase space, enabling on-the-fly estimation of the simulation accuracy and offering a practical alternative to the often costly verification of Hamiltonian conservation.

\section*{Data availability}

The propagation schemes we use for the ${\bf E}\times{\bf B}$ and KS TDDFT simulations are publicly available on their respective repositories~\cite{GC2D,QMol_grid}. 

\begin{acknowledgments}
CC acknowledges useful discussions with J. Dubois. This work has been carried out within the framework of the EUROfusion Consortium, funded by the European Union via the Euratom Research and Training Programme (Grant Agreement No 101052200 — EUROfusion). Views and opinions expressed are however those of the author(s) only and do not necessarily reflect those of the European Union or the European Commission. Neither the European Union nor the European Commission can be held responsible for them. This material is based upon work supported by the National Science Foundation under Grant No.~PHY-2207656.
\end{acknowledgments}

\appendix

\section{Equations to implement for TDDFT extended phase-space split-operators} \label{app:TDDFT_split_operator_equations}

In this appendix, we summarize the equations that need to be implemented for extended phase-space split-operator propagators with TDDFT. 
Recall that the extended phase-space propagation involves pairs of KS orbitals $\phi$ and $\overline{\phi}$ from which all the split-operator components are calculated -- see the discussion at the beginning of section~\ref{sec:TDDFT:simulation}.
For simplicity, we write the propagation steps that evolve $\phi$ and the ones that evolve $\overline{\phi}$ are trivially obtained by swapping $\phi\leftrightarrow\overline{\phi}$ in these equations.
When a phase space variable is omitted in the list below, it means that the corresponding flow $\text{e}^{\tau\mathcal{L}_{H_\text{el}}}$ leaves that variable unchanged.
As in the main text, we use atomic units such that the reduced Plank constant and charge of the electron are both equal to 1 and omitted in the equations.

\textbf{Kinetic term} for field-free or (dipole approximation) length-gauge dynamics:
$$
    \text{e}^{\tau \mathcal{L}_{H_\text{kin}[\phi]}}\phi \to 
        \text{e}^{i \tau \frac{\Delta}{2}}\phi = 
        \mathcal{F}^{-1}\left[\text{e}^{-i \tau \frac{{\bf p}^2}{2}}\mathcal{F}[\phi]\right],
$$
where the last part of the equation can be used to calculate the kinetic contributions with fast Fourier transforms (as in \texttt{QMol-grid}).

\textbf{Kinetic term} for (dipole approximation) velocity-gauge dynamics:
First, because the kinetic component contains an explicit time dependence (via the vector potential), when tracking the conservation of energy the variable $\xi$ must be updated
$$
    \text{e}^{\tau \mathcal{L}_{H_\text{kin}[\phi]}}\xi \to 
        \xi + \tau {\bf E}(t) \cdot \sum_k{n_k\int{
                \phi_k^\dagger (-i\nabla+{\bf A}(t)) \phi_k\ \text{d}{\bf x}
            }},
$$
where ${\bf E}$ and ${\bf A}$ are the electric field and vector potential, respectively.
Then, the KS orbitals are updated with
$$
    \text{e}^{\tau \mathcal{L}_{H_\text{kin}[\phi]}}\phi \to 
        \text{e}^{-i \tau \frac{(-i\nabla+{\bf A}(t)^2}{2}}\phi = 
        \mathcal{F}^{-1}\left[\text{e}^{-i \tau \frac{({\bf p}+{\bf A}(t)^2}{2}}\mathcal{F}[\phi]\right].
$$

\textbf{Potential term} for the explicit part of the potential (\textit{i.e.}, with potential split) -- see main text:
First the potential $\hat{V}_\text{exp}=\hat{V}[\phi]$ must be initialized, which includes the calculation of the one-body density $\rho(\phi)$. If the system is externally driven and in the length gauge, the dipole contribution must be added to the potential
$$
    \hat{V}[\phi] \to \hat{V}[\phi] + {\bf E}(t)\cdot {\bf x}
$$
and, when tracking energy conservation, the variable $\xi$ must be updated (again, because of the explicit time dependence in the potential component)
$$
    \text{e}^{\tau \mathcal{L}_{H_\text{exp}[\phi]}} \xi \to
        \xi - \tau\dot{\bf E}(t)\cdot\int{{\bf x}\ \rho(\phi)\ \text{d}{\bf x}}.
$$
Then, the KS orbitals are updated with
$$
    \text{e}^{\tau \mathcal{L}_{H_\text{exp}[\phi]}}\phi \to 
        \text{e}^{-i \tau \hat{V}[\phi]}\phi.
$$

\textbf{Potential term} without potential split or for the implicit part of the potential -- see main text:
Like for the explicit-potential part, the first step is to initialize the full $\hat{V}$ or implicit part $\hat{V}_\text{imp}$ of the potential, which we both note $\hat{V}[\overline{\phi}]$ in what follows. Note that here the initialization must be performed with the other set of KS orbitals $\overline{\phi}$.

Without potential split and when the system is externally driven in the length gauge, the dipole contribution and variable $\xi$ (when tracking the energy) are respectively updated as
$$
    \hat{V}[\phi] \to \hat{V}[\phi] + {\bf E}(t)\cdot {\bf x},
$$
and
$$
    \text{e}^{\tau \mathcal{L}_{H_\text{pot/imp}(\phi)}} \xi \to
        \xi - \tau\dot{\bf E}(t)\cdot\int{{\bf x}\ \rho(\overline{\phi})\ \text{d}{\bf x}}.
$$
Note that here the one-body density associated with the $\overline{\phi}$ KS orbitals should be used in calculating $\xi$, for consistency with the definition of $H_\text{pot}$ and $H_\text{imp}$.

Finally, the KS orbitals are updated with
$$
    \text{e}^{\tau \mathcal{L}_{H_\text{pot/imp}[\phi]}}\phi \to 
        \phi -i \tau \hat{V}[\overline{\phi}] \phi.
$$

\textbf{Restraint}: The restraint updates both $\phi$ and $\overline{\phi}$ at the same time and thus requires a local copy of the KS orbitals -- in \texttt{QMol-grid} we use three sets of KS orbitals: one for $\phi$, one for $\overline{\phi}$, and one local copy to evaluate the restraint and when we need the average $(\phi+\overline{\phi})/2$. 
Following Eqs.~\eqref{eq:TDDFT_restrain_flow}, the KS orbitals are updated with
$$
    \text{e}^{\tau {\mathcal{L}}_{R+\xi}}\phi \to 
        \frac{\phi+\overline{\phi}}{2} + 
        \text{e}^{-2 i \omega \tau} \frac{\phi-\overline{\phi}}{2},
$$
and
$$
    \text{e}^{\tau {\mathcal{L}}_{R+\xi}}\overline{\phi} \to 
        \frac{\phi+\overline{\phi}}{2} - 
        \text{e}^{-2 i \omega \tau} \frac{\phi-\overline{\phi}}{2}.
$$

As alluded in the notation above recall that, for externally driven systems, we group the $\xi$ component of the Hamiltonian functional with the restraint. This leads to the intermediate time update
$$
    \text{e}^{\tau {\mathcal{L}}_{R+\xi}} t \to t + \tau.
$$

\section{Test system for TDDFT propagators} \label{app:TDDFT_test_system}

We evaluate the accuracy and efficacy of the propagators by performing driven time-dependent spin-restricted Hartree-Fock calculations of far-from-equilibrium dynamics, which corresponds to TDDFT using the exact exchange and no correlation functionals in the DFT Hamiltonian. 
Specifically, considering a one-dimensional molecular model very similar to that of Ref.~\cite{Mauger_2023}, we systematically calculate the (i) error in the final KS orbitals after 150-a.u.\ propagation, (ii) wall time for propagating over the first 60~a.u., and (iii) distance between the extended phase-space copies between 60 and 150~a.u. 
We then evaluate the error by taking the $L^2$ norm of the difference between the final Kohn-Sham orbitals and that obtained with a time step of $5\times10^{-3}$ (twice as small as the smaller time step we report on the figure).
In Fig.~\ref{fig:TDDFT_compare_restrains}~(b), we report the $L^2$ norm of the distance~\eqref{eq:TDDFT_extended_phase_space_distance} 
$
    \sqrt{\int_{60}^{150}{d(t)^2\ {\rm d}t}}
$.

We have checked that we obtain similar results for (i) field-free and field-driven dynamics, (ii) spin restricted and spin polarized models, and (iii) other common observables such as the error in the final one-body density and time-dependent dipole signal.

\section{Operator split scheme} \label{app:split_scheme}

\begin{widetext}
We have implemented four different extended phase-space split-operator schemes in \texttt{QMol-grid}~\cite{mauger_2024,QMol_grid}, respectively defined by the expansion coefficients in Eq.~\eqref{eq:symplectic_split_operator}:
\begin{itemize}
    \item 2$^\text{nd}$-order Strang ({\it a.k.a.} Verlet, $p=2$ and $K=1$)~\cite{Strang_1968}: ${\bf a}=\{0.5\}$.
    \item 4$^\text{th}$-order Forest-Ruth ($p=4$ and $K=3$)~\cite{Forest_1990}: 
        ${\bf a}=\{ \frac{1}{2(2 - 2^{1/3})}, \frac{1}{2(2 - 2^{1/3})}, 
            \frac{1}{2}-\frac{1}{2(2 - 2^{1/3})} \}$.
    \item optimized 4$^\text{th}$-order Blanes and Moan ($p=4$ and $K=6$)~\cite{Blanes_2002}:
        ${\bf a}=\{$0.0792036964311957, 0.1303114101821663, 0.2228614958676077,
                    \mbox{-0.3667132690474257}, 0.3246481886897062, 0.1096884778767498$\}$.
                    
    \item optimized 6$^\text{th}$-order Blanes and Moan ($p=6$ and $K=10$)~\cite{Blanes_2002}:
        ${\bf a}=\{$0.050262764400392, 0.098553683500650, 0.314960616927694,
                    \mbox{-0.447346482695478}, 0.492426372489876, \mbox{-0.425118767797691},
                    0.237063913978122, 0.195602488600053, 0.346358189850727, 
                    \mbox{-0.362762779254345}$\}$.
\end{itemize}
\end{widetext}
Other split schemes~\cite{Yoshida_1990,McLachlan1995,Omelyan2002,Blanes_2002,Blanes_2013} are also available in the Python package \texttt{pyhamsys}~\cite{pyhamsys}.

\begin{figure}[htb]
    \centering
    \includegraphics[width=\linewidth]{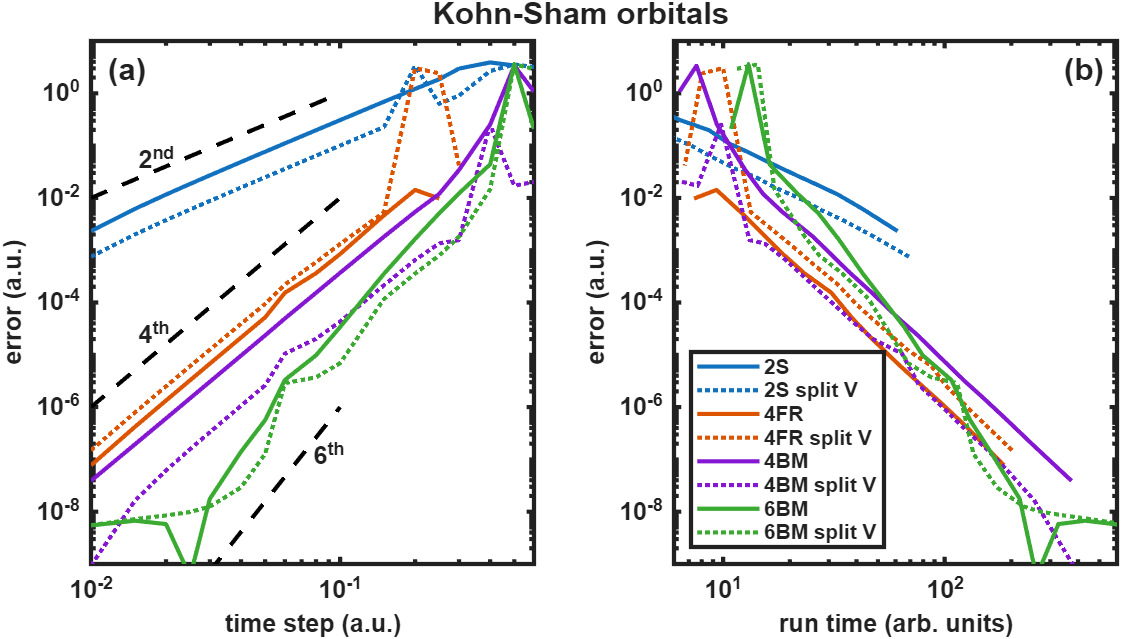}
    \caption{\label{fig:TDDFT_compare_schemes} 
    Comparison of the (a) accuracy and (b) efficacy of extended phase-space symplectic integration using different split-operator schemes -- see legend and main text. 
    2S corresponds to the 2$^\text{nd}$-order Strang ({\it a.k.a.} Verlet) scheme, 4FR to the 4$^\text{th}$-order Forest-Ruth, and 4/6BM to the optimized 4/6$^\text{th}$-order Blanes and Moan.
    For all calculations, we use a restraint coefficient $\omega=10$ and TRV split motif. We attribute the plateauing of the accurary/efficiency for the 6BM results at the smallest time steps to reaching machine precision in the propagation~\cite{Mauger_2023}.
    For reference, the dashed black lines in panel (a) mark perfect 2$^\text{nd}$-, 4$^\text{th}$-, and  6$^\text{th}$-order scalings.}
\end{figure}

Figure~\ref{fig:TDDFT_compare_schemes} compares the (a) accuracy and (b) efficacy for the four split-operator schemes in \texttt{Qmol-grid} listed above.
Similar to the discussion in the main text, in panel (a) we see that the convergence rate of all schemes match their order and that splitting the potential generally leads to improved accuracy, to the exception of the 4$^\text{th}$ order Forest-Ruth where it slightly worsen it. Unsurprisingly, we also see that increasing the order and using optimized schemes notably improves the results.
The efficacy of panel~(b) shows that, for acceptable level of error in simulations, the 2$^\text{th}$ Strang ({\it a.k.a.} Verlet; 2S curves) is never optimal. Instead, for the test calculations we consider here either the 4$^\text{th}$-order Forest-Ruth (4FR) without potential split or optimized Blanes-Moan (4BM) with potential split schemes are the most efficient. Finally, the optimized 6$^\text{th}$-order optimized Blanes and Moan (6BM) schemes essentially reaches machine precision before it can overtake the most efficient 4$^\text{th}$-order schemes -- see the plateauing of the accuracy/efficacy at the smallest time steps. Overall, these results are consistent with what we observed in cases where the DFT functional is fully explicit and no phase-space extension is required~\cite{Mauger_2023}.

\section{Comparison of restraint and midpoint projection in TDDFT simulations} \label{app:restraint_vs_MPP}

\begin{figure}[htb]
    \centering
    \includegraphics[width=\linewidth]{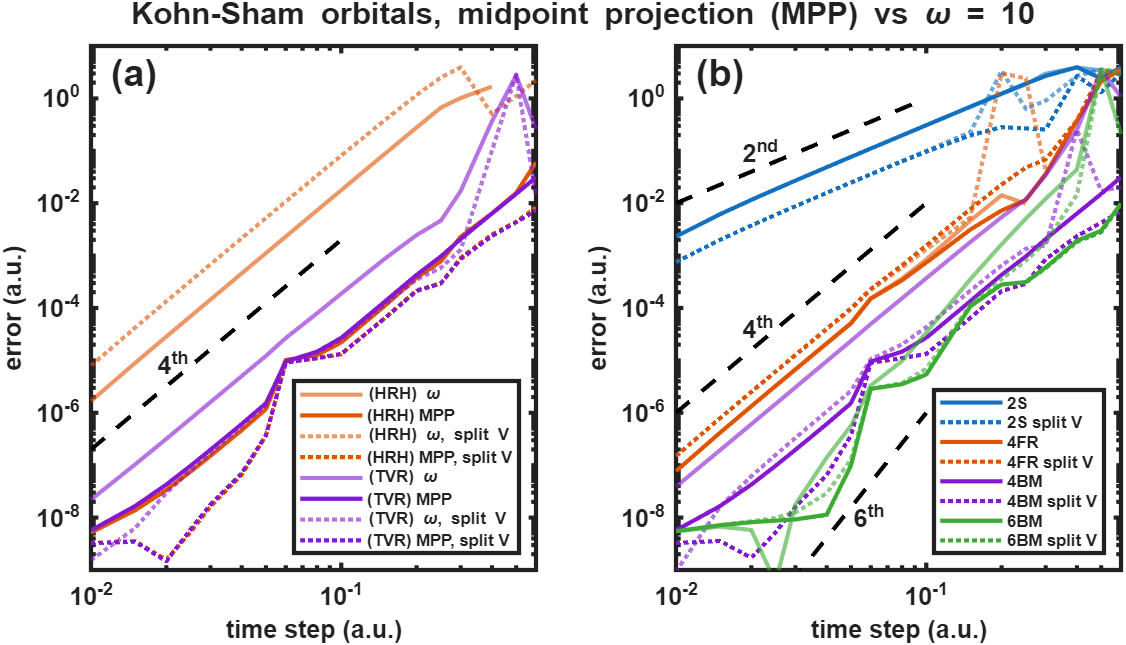}
    \caption{\label{fig:TDDFT_restraint_vs_MPP} 
    Comparison of the accuracy of the restraint with $\omega=10$ and midpoint projection (MPP) in TDDFT simulations for (a) different types and order of terms in the optimized 4$^\text{th}$-order Blanes and Moan, and (b) different schemes.
    In both panels, full-color solid and dotted curves correspond to the midpoint projection while the matching-color partially transparent ones are for the restraint and reproduced from figure~\ref{fig:TDDFT_compare_splits}~(a) and figure~\ref{fig:TDDFT_compare_schemes}~(b), respectively.
    For reference, the dashed black lines mark perfect n$^\text{th}$-order scaling, with n marked next to each line.
    }
\end{figure}

Figure~\ref{fig:TDDFT_restraint_vs_MPP}~(a) compares the accuracy of the midpoint projection and $\omega=10$ restraint for the worse (HRH) and best (TVR) splitting order of figure~\ref{fig:TDDFT_compare_splits}~(a) -- compare the pairs of curves with matching colors and style but different transparency. Unlike the restraint case, the midpoint projection results are relatively insensitive to the order of the terms in the split operator, and splitting the potential yields less substantial improvement as compared to the case where the implicit and explicit parts are kept together. 
Figure~\ref{fig:TDDFT_restraint_vs_MPP}~(a) performs a similar comparison but for different split operator schemes. In this case, we see that only the optimized 4/6$^\text{th}$-order Blanes and Moan schemes yield a notable difference between the restraint and midpoint projection. On the other hand, the 2$^\text{nd}$-order Strang and 4$^\text{th}$-order Forest-Ruth results are essentially on top of each other, except for some of the larger time steps where the restraint likely encounters some resonances.
When splitting the potential and the (TVR) split of panel~(a) as well as all the split-potential cases of panel~(b), taking the restraint $\omega=0$ without the midpoint projection for which figure~\ref{fig:TDDFT_compare_restrains}~(b) shows the highest accuracy, we observe essentially no difference with the midpoint projection (not shown). The (HRH) split also matches the midpoint projection for time steps smaller than 0.1~a.u.

%

\end{document}